\RequirePackage{fix-cm}
\documentclass[smallextended]{svjour3}       
\smartqed  
\usepackage{graphicx}
\usepackage{natbib}
\usepackage{txfonts}
\begin{document}

\title{Planet formation in Binaries}

\titlerunning{Planet formation in Binaries}        

\author{P. Thebault \and N. Haghighipour}


\institute{P. Thebault \at
LESIA, Observatoire de Paris,
F-92195 Meudon Principal Cedex, France \\
              \email{philippe.thebault@obspm.fr}           
\and
N.Haghighipour \at
Institute for Astronomy and NASA Astrobiology Institute, University of Hawaii-Manoa, 2680 Woodlawn Drive Honolulu, HI 96822 USA
}


\maketitle

\begin{abstract}

Spurred by the discovery of more than 60 exoplanets in multiple systems, binaries have become in recent years one of the main topics in planet formation research. Numerous studies have investigated to what extent the presence of a stellar companion can affect the planet formation process. Such studies have implications that can reach beyond the sole context of binaries, as they allow to test certain aspects of the planet formation scenario by submitting them to extreme environments. 
We review here the current understanding on this complex problem. We show in particular how each of the different stages of the planet-formation process is affected differently by binary perturbations. We focus especially on the intermediate stage of kilometre-sized planetesimal accretion, which has proven to be the most sensitive to binarity and for which the presence of some exoplanets observed in tight binaries is difficult to explain by in-situ formation following the "standard" planet-formation scenario. Some tentative solutions to this apparent paradox are presented.
The last part of our review presents a thorough description of the problem of planet habitability, for which the binary environment creates a complex situation because of the presence of two irradiation sources of varying distance.

\end{abstract}

\section{Introduction} \label{intro}

About half of solar-type stars reside in multiple stellar systems \citep{ragh10}. As a consequence, one of the most generic environments to be considered for studying planet formation should in principle be that of a binary. However, the "standard" scenario of planet formation by core-accretion that has been developed over the past decades \citep[e.g.][]{safr72,liss93,poll96,hubi05} is so far restricted to the case of a single star. Of course, this bias is a direct consequence of the fact that planet formation theories were initially designed to understand the formation of our own solar system. For the most part, however, this bias is still present today, nearly 2 decades after the discovery of the first exoplanets. The important updates and revisions of the standard model, such as planetary migration, planet scattering, etc., that have been developed as a consequence of exoplanet discoveries have mostly been investigated for a single star environment. This single-star-environment tropism does also affect the alternative planet-forming scenario by gravitational instabilities \citep{boss97}, which has witnessed a renewed interest after the discovery of jovian exoplanets at large radial distances from their star \citep{boss11}.

However, the gradual discovery of exoplanets in multiple star systems \citep[][ and references therein]{desi07,mugr09}, and especially in close-binaries of separation $\sim 20\,$AU, has triggered the arrival of studies investigating how such planets could come about and, more generally, about how planet-formation is affected by binarity. The latter issue is a vast and difficult one. Planet formation is indeed a complex process, believed to be the succession of several stages \citep[e.g.][]{hagh11}, each of which could be affected in very different ways by the perturbations of a secondary star. 

Not surprisingly, the effect of binarity on each of these different stages is usually investigated in separate studies. 
A majority of these investigations have focused on the intermediate stage of kilometre-sized planetesimal accretion, as this stage has been shown to be extremely sensitive to stellar companion perturbations. But other key stages have also been explored, from the initial formation of protoplanetary discs to the final evolution of massive planetary embryos.

The scope of such planets-in-binaries studies has been recently broadened by the discovery of several \emph{circumbinary} exoplanets (also known as P-type orbits) by the Kepler space telescope, most of which are located relatively close to the central stellar couple. The issues related to the formation of these objects are often very different from those related to circumprimary exoplanets (also known as S-type orbits), and in-depth investigations of this issue have only just begun.

Studying how both circumprimary and circumbinary planets form is of great interest, not only to explore the history of specific planet-hosting binaries, but also for our understanding of planet formation in general. 
These studies can indeed be used as a test bench for planet formation models, by confronting them to an unusual and sometimes "extreme" environment where some crucial parameters might be pushed to extreme values.
We present here a review of the current state of research on planets in binary star systems. We focus our analysis on the S-type systems in which planets orbit one star of the binary.

\section{Observational constraints: Planets in binaries}

\begin{figure}
\includegraphics[width=\columnwidth]{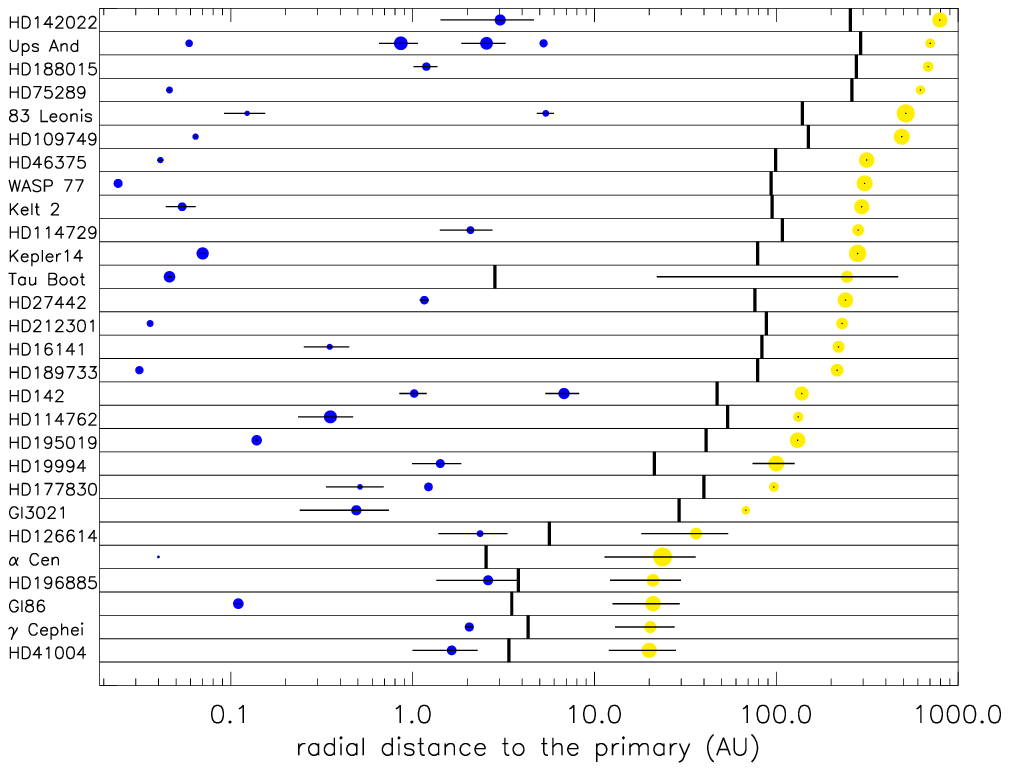}
\caption[]{Architecture of all circumprimary planet-bearing binaries with separation $\leq 1000\,$AU (as of July 2013). Companion stars are displayed as yellow circles, whose radius is proportional to $(M_2/M_1)^{1/3}$. Planets are marked as blue circles whose radius is proportional to $(m_{pl}/m_{Jup})^{1/3}$. The horizontal lines represent the radial excursion of the planets and stars orbit (when they are known). For most binaries of separation $\geq 100\,$AU, the orbit is not known and the displayed value corresponds to the projected current separation. The short horizontal lines correspond to the outer limit of the orbital stability region around the primary, as estimated by \citet{holw99}.
}
\label{binarch}
\end{figure}

Exoplanet search surveys were initially strongly biased against binary systems of separation $\lesssim 200\,$AU \citep{egge10}, in great part because these searches were focusing on stellar environments as similar as possible to the solar system. 
In 2003, however, the first exoplanet in a close binary was detected in the $\gamma$ Cephei system \citep{hatz03}, and today more than 60 exoplanets are known to inhabit multiple star systems \citep{roel12}. Note that, in many cases, these exoplanets were detected $before$ the presence of a stellar companion was later established by imaging campaigns \citep{mugr09}. As a result, for most of these systems, the separation of the binary is indeed relatively large, often in excess of 500\,AU \citep{roel12}. However, $\sim 10$ of these planet-bearing binaries have a separation of less than 100\,AU, with 5 exoplanets in close binaries with separations of $\sim 20\,$AU (Fig.\ref{binarch}): Gl86 \citep{quel00,lag06}, HD 41004 \citep{zuck04}, $\gamma$ Cephei \citep{hatz03,neuh07,endl11}, HD196885 \citep{corr08,chau10} and $\alpha$ Centauri B \citep{dumu12}.

As soon as their number became statistically significant, the characteristics of these planets in binaries have been investigated in order to derive possible specificities as compared to planets around single stars. \citet{desi07} and \citet{roel12} have shown that, while the distribution of planets in wide ($\geq 100\,$AU) systems is identical to that of planets around single stars, the characteristics of exoplanets in close binaries are significantly different. The main trend seems to be that planetary masses increase with decreasing stellar separation. According to \citet{roel12}, the minimum planet mass scales approximately as $(10\rm{AU}/a_{bin})M_{Jup}$. However, these trends should be taken with caution, as the number of planets in tight binaries is still very limited. Furthermore, the tentative detection of the Earth-sized planet around $\alpha$ Cen B in late 2012 might significantly weaken this result.
As for the global occurrence of planets in binaries, \citet{roel12} have found that multiplicity rate among planet-hosting stars is $\sim 12\%$, approximately four times smaller than for main field solar type stars \citep{ragh10}. But as pointed out by \citet{duch10}: "...the small sample size, adverse selection biases, and incompleteness of current multiplicity surveys are such that it is premature to reach definitive conclusions". As a consequence, future surveys should probably increase both the number of exoplanets in close-binaries, and the number of stellar companions in known exoplanet systems.

Besides these statistical exploration, another crucial issue that has been investigated early on is that of the long term orbital stability of these planets in binaries. The reference work on this issue remains probably that of \citet{holw99} \footnote{Although similar pioneering work on this issue had already been performed a decade earlier by \citet{dvo84,dvo86} and \citet{dvo89}}, who derived empirical expressions for orbital stability as a function of binary semi-major axis $a_B$, eccentricity $e_B$ and mass ratio $\mu$. Later studies have shown that, reassuringly, all known exoplanets in multiple systems are on stable orbits \citep[e.g.][]{dvo03,hagh10}, although the case for HD41004 is not fully settled yet, as it depends on the yet unconstrained eccentricity of the binary orbit \citep{hagh10}.

An important recent development in terms of observations is the discovery of several exoplanets in P-type orbits. The first confirmed such planet orbits around the cataclysmic binary DP Leonis \citep{qian10}, but most circumbinary planets around binaries with main sequence stars have been detected by the Kepler space telescope \citep{doyl11,wels12,oros12a,oros12b, schw13,kost13,kost14a,kost14b}. Here again, dynamical studies have shown that all known circumbinary planets are on long term stable orbits. 

However, even if the question of long term stability seems to be settled for all know exoplanets in binaries, the question of their \emph{formation} is a much more complex issue. It is true that, for many S-type planets, the stellar separation is so large, often exceeding 100 times the radial distance of the exoplanet to the primary, binarity should have had a very limited effect in the planet-forming regions. The situation should, however, be radically different for the handful of planets in $\sim 20\,$AU binaries, notably for $\gamma$ Cephei Ab, HD196885 Ab and HD 41004 Ab, where the planet is located close to the orbital stability limit (Fig.\ref{binarch}). It is unlikely that planet formation in these highly perturbed environments could proceed unaffected by the presence of the companion star.

\section{Early Stages of Planet Formation} \label{early}

\subsection{Protoplanetary Disc Truncation}

\begin{figure}
\makebox[\textwidth]{
\includegraphics[width=.5\columnwidth]{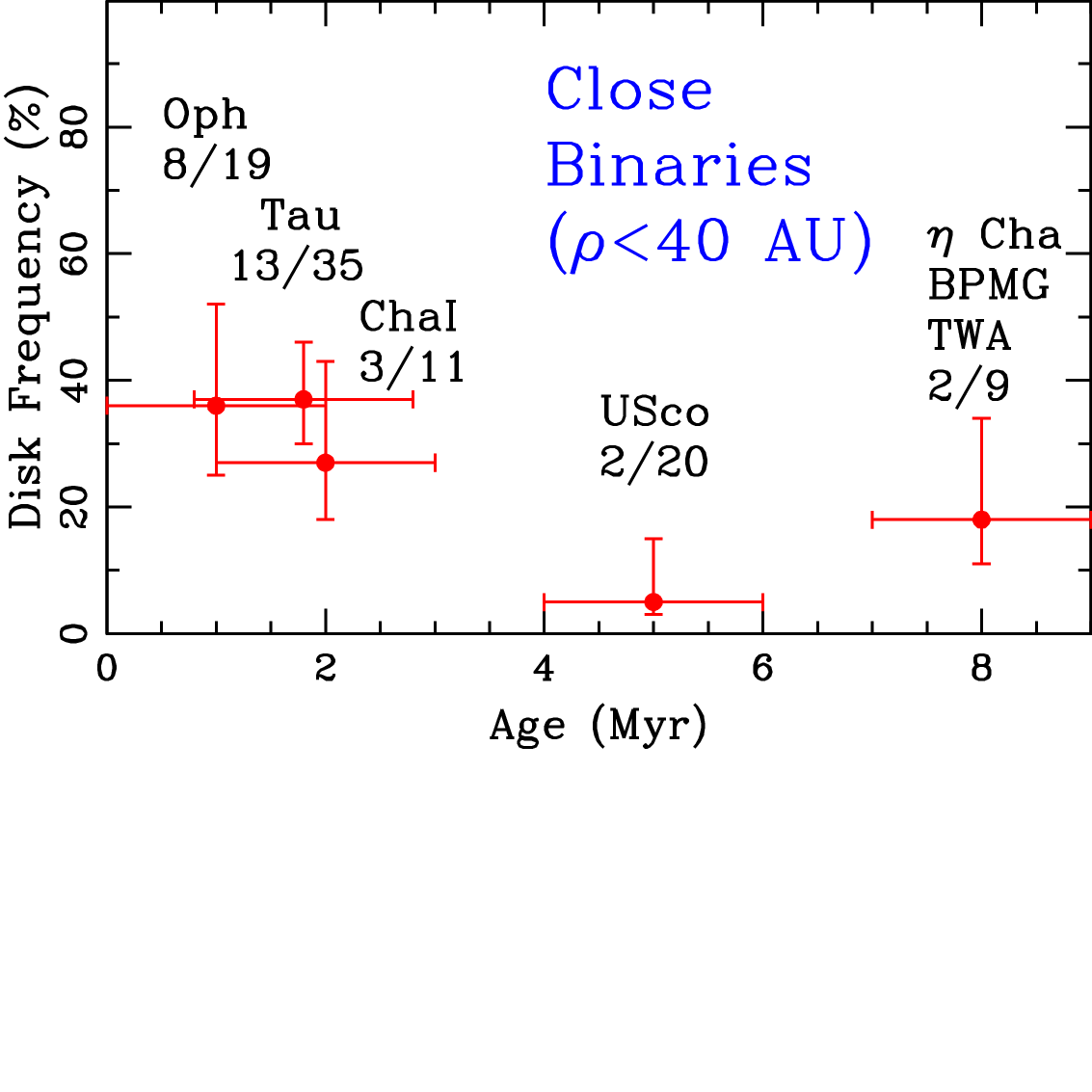}
\hfil
\includegraphics[width=.5\columnwidth]{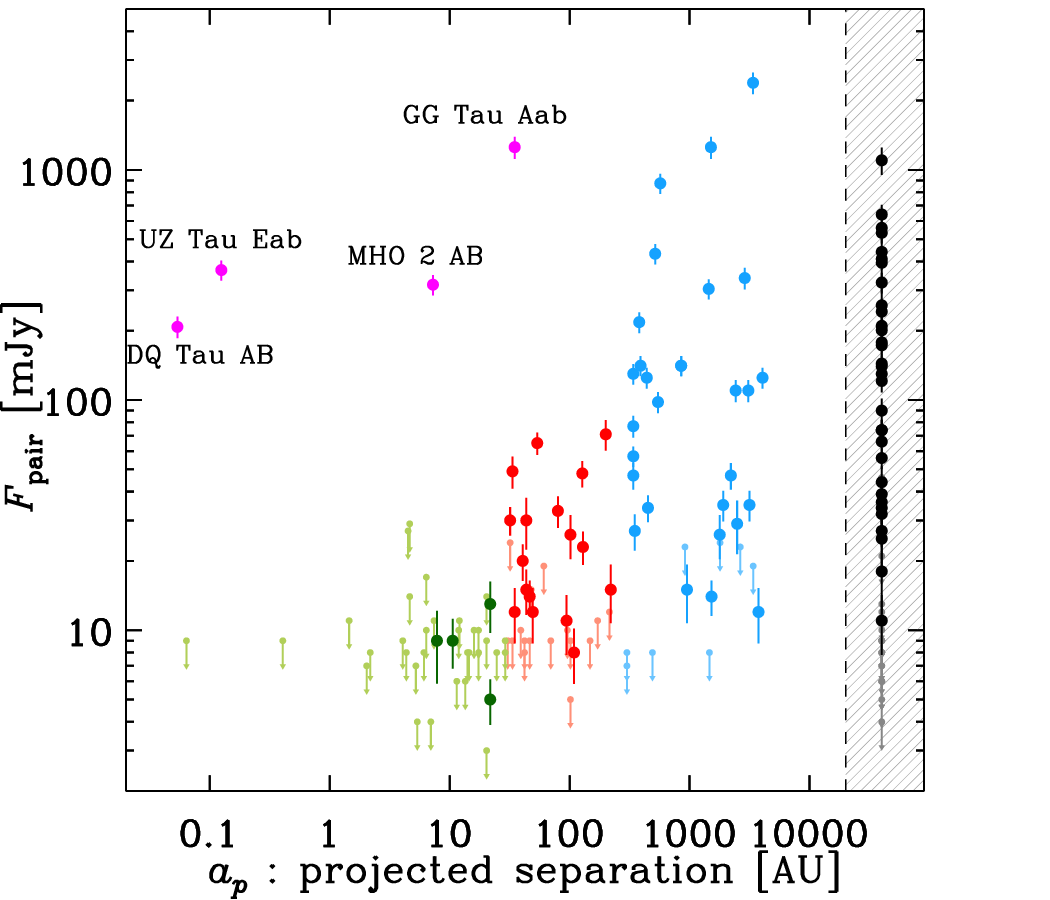}
}
\caption[]{Observational contraints on discs in binaries. \emph{left panel}: Frequency of disc-bearing systems in close binaries as a function of age in several young stellar clusters \citep[from][ courtesy of the Astrophysical Journal]{krau12}. \emph{right panel}: Measured excess flux in the millimetre as a function of binary projected separation  \citep[from][ courtesy of the Astrophysical Journal]{harr12}
}
\label{harriskraus}
\end{figure}

The planet formation process can be affected by binarity right from the start, during the formation of the initial massive gaseous protoplanetary disc. \citet{arty94} and \citet{savo94} have shown that such a disc can be tidally truncated by a stellar companion. This truncation distance depends on several parameters, such as the disc's viscosity, but is in most cases roughly comparable to the outer limit for dynamical stability (see previous section), i.e., typically at 1/3 to 1/4 of the binary's separation for non-extreme value of the orbital eccentricity.

Surveys of discs around young stellar objects (YSOs) have given observational confirmation that discs in close binaries indeed tend to be both less frequent and less massive than around single stars. For the 2Myr old Taurus-Auriga association, \citet{krau12} have shown that, while the disc frequency in wide binaries remains comparable to that of single stars ($\sim 80$\%), it significantly drops for separations $\lesssim 40\,$AU and is as low as $\sim 35$\% for binaries tighter than 10\,AU (Fig.\ref{harriskraus}a). Moreover, the millimetre-wave dust continuum imaging survey of \citet{harr12} for discs in the same Taurus-Auriga cluster have shown that the estimated dust mass contained in these discs strongly decreases with decreasing binary separation (Fig.\ref{harriskraus}b). This agrees very well with the theoretical prediction of more compact, and thus less massive discs in tight binaries.

This less-frequent and more-tenuous discs trend leads to two major problems when considering planet formation. This first one is that truncated discs might not contain enough mass to form planets, especially Jovian objects such as those that have been observed for the three most "extreme" systems: HD196885, HD 41004 and $\gamma$ Cephei. This issue has been numerically investigated for the specific case of $\gamma$ Ceph., for which \citet{jang08} have found that, for the most reasonable assumptions regarding the disc's viscosity and accretion rate, there is just enough mass left in the truncated disc to form the observed giant planet. This encouraging result has been later confirmed by \citet{mull12} using a different numerical approach. Note, however, that there might not be enough mass left in the truncated $\gamma$ Ceph disc to form another Jovian planet.

There is, however, a potentially bigger problem that is inherent to truncated discs, i.e., that they should be short-lived. The viscous evolution of a compact disc is indeed much faster than that of an extended system, and its mass gets drained, by accretion onto the central star, on shorter timescales. This reduces the timespan within which gaseous planet can form. As an example, \citet{mull12} have shown that, for $\gamma$ Cephei type systems, only for unrealistically low disc viscosities do they obtain disc lifetimes that are long enough to allow for in-situ giant planet formation by core-accretion.
Observational confirmation of this shortlived-discs trend has been obtained by \citet{krau12}, who compared the disc frequency as a function of age for close ($\leq 40\,$AU) binaries in several nearby young associations. They found that the majority of such systems lose their disc in less than 1 Myr, even if a small fraction is able to retain discs to ages close to 10Myr. Their preliminary conclusion is that "$\sim 2/3$ of all close binary systems clear their disks extremely quickly, within 1 Myr of the end of envelope accretion. The other $\sim 1/3$ of close binary systems evolve on a timescale similar to that of single stars". There seems thus to exist a disc-in-binary category for which the theoretically-expected faster mass drain does not occur, and which could thus be more friendly to planet formation. A good example for this category could be the young L1551 system, harboring two resolved $\sim10\,$AU wide circumstellar disks in a binary of $45\,$AU separation \citep{rodr98}. Note, however, that even for this population of long-lived binary discs, most estimated disc masses are much lower than for single star cases (with, however, some important exceptions, such as L1551).

\subsection{Grain Condensation and Growth}

The next stage of planet formation, the condensation of small grains and their growth into larger pebbles and eventually kilometre-sized planetesimals, has not been extensively studied in the context of binary systems. One main reason is probably that this stage is the one that is currently the least understood even in the "normal" context of single stars \citep[e.g.][]{blum08}, so that extrapolating it to perturbed binaries might seem premature. A noteworthy exception is the study by \citet{nels00} showing that for an equal-mass binary of separation 50\,AU, temperatures in the disc might stay too high to allow grains to condense. These results were recently confirmed by the sophisticated 3-D modelling of radiative discs by \citet{pico13}, who found strong disc heating, due of spiral chocks and mass streaming between the circumprimary and circumsecondary discs, in binaries of separation 30 and 50\,AU.
On a related note, \citet{zsom11} showed that, even if grains can condense, binary perturbations might impend their growth by mutual sticking because of too high impact velocities.

\section{Planetesimal Accumulation} \label{middle}

\subsection{Context}

The next stage of planet formation starts once "planetesimals", i.e., objects large enough (typically sub-kilometre to kilometre-sized) to decouple from the gas, have formed. Within the standard planet-formation scenario, this stage leads, by mutual accretion among these km-sized planetesimals, to the formation of large planetary embryos. 
This stage is the one that has been by far the most extensively studied within the context of binary systems. The main reason is that the accretion of planetesimals and their growth to planetary embryos is potentially extremely sensitive to dynamical perturbations, since it does not take much to hamper or even halt the mutual accretion of kilometre-sized objects. Indeed, in the standard simulations of planet-formation around single stars, this stage proceeds through fast runaway and oligarchic growth that require very low impact velocities between colliding bodies, typically smaller than their escape velocity, i.e., just a few m.s$^{-1}$ for kilometre-sized objects \citep[e.g.,][]{liss93,koku00}. These values are less than $\sim 10^{-4}$ that of typical orbital velocities, so that even moderate dynamical perturbations can have a dramatic effect by increasing impact velocities beyond an accretion-hostile threshold.

The crucial parameter sealing the fate of the planetesimal swarm is thus their impact velocities. As a consequence,  most studies exploring this stage have numerically investigated how encounter velocities evolve under the coupled effect of stellar perturbations, gas drag and physical collisions.

\subsection{Differential Orbital Phasing: the negative effect of gas drag}


In a co-planar system, when neglecting all other effects, the response of a planetesimal disc to the pure gravitational perturbations of a stellar companion are secular perturbations that lead to oscillations of their eccentricity and longitude of periastron $\omega$ \citep{hepp78,theb06,giup11}. These secular oscillations are strongly phased in $\omega$, so that they initially do not lead to large impact velocities $\Delta v$ for planetesimal encounters. However, as these oscillations get narrower with time, at some point they are narrow enough for orbital crossing to occur, leading to a sudden increase of $\Delta v$ \citep[see][for a detailed discussion on this issue]{theb06}.

\begin{figure*}
\makebox[\textwidth]{
\includegraphics[width=.5\columnwidth]{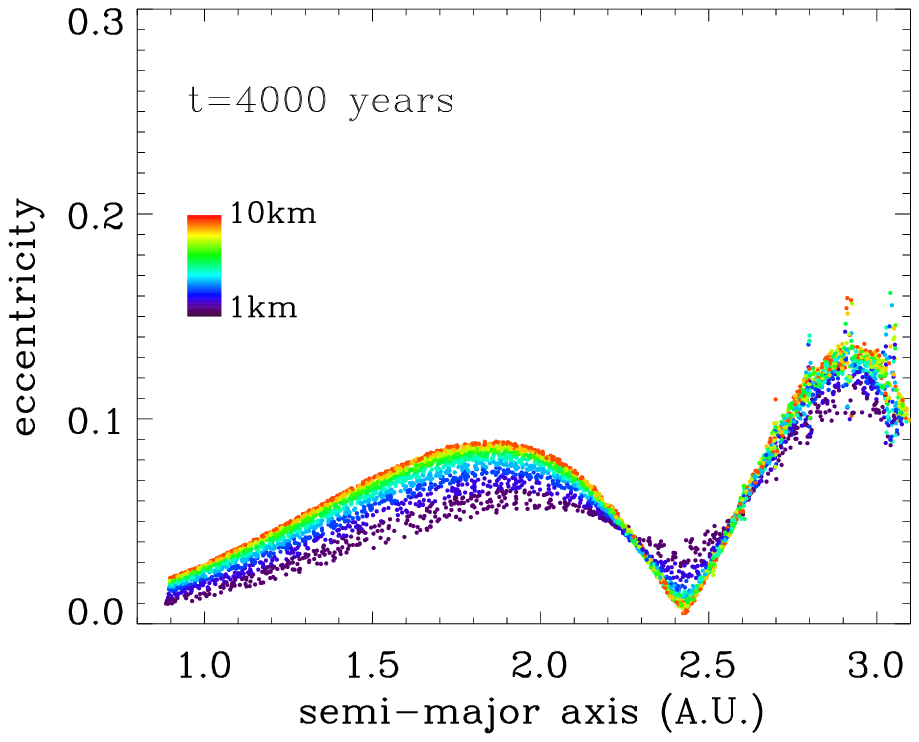}
\hfil
\includegraphics[width=.5\columnwidth]{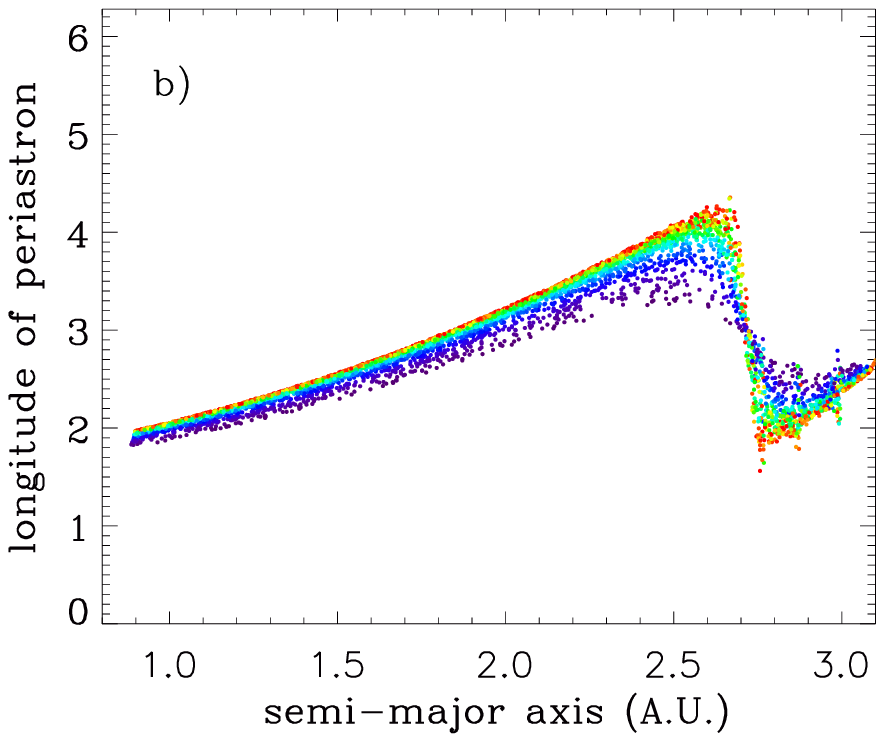}
}
\vspace*{-40pt}
\makebox[\textwidth]{
\includegraphics[width=.5\columnwidth]{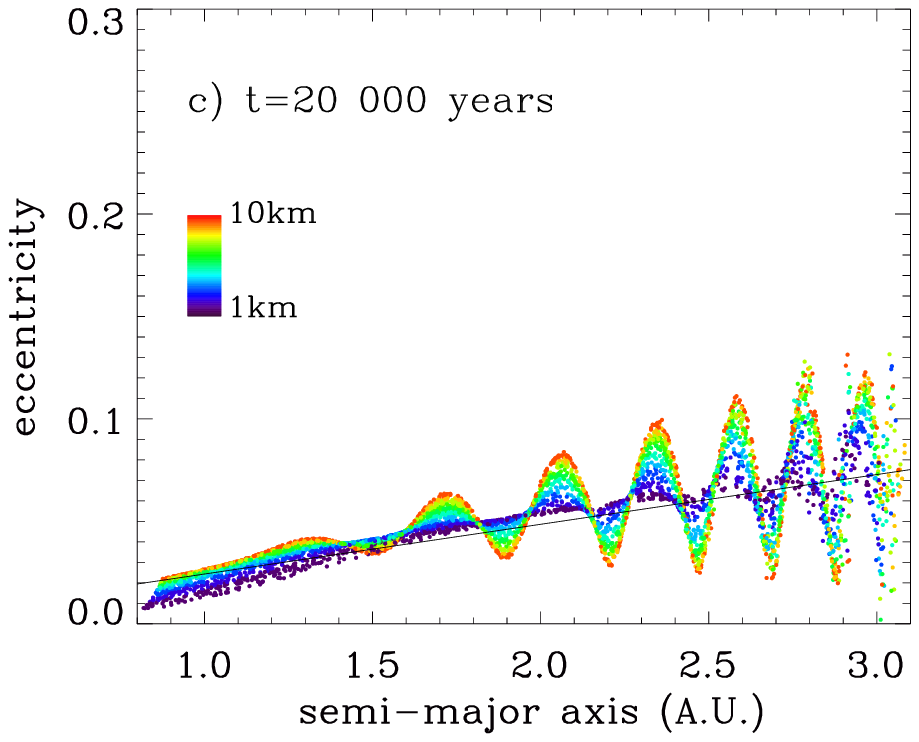}
\hfil
\includegraphics[width=.5\columnwidth]{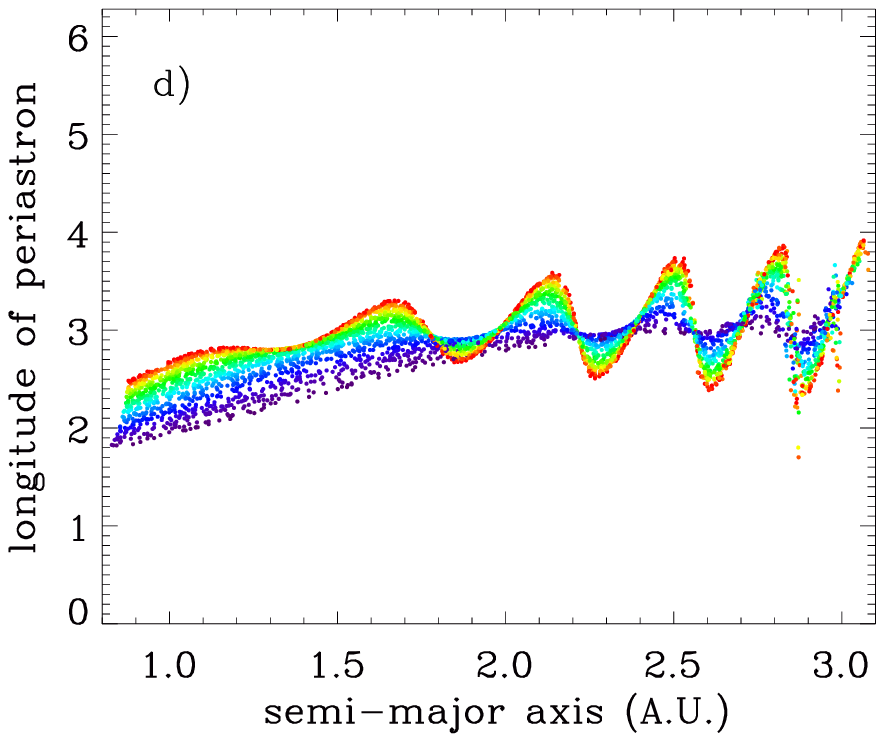}
}
\vspace*{20pt}
\caption[]{Evolution of a planetesimal population in a close binary when taking into account gas drag. \emph{Left panel:} eccentricity. \emph{Right panel}: longitude of periastron (the binary's longitude of periastron is 0). The binary configuration corresponds to the HD196885 case, i.e., $M_2/M_1=0.35$, $a_b=21\,$AU and $e_b=0.42$. The gas disc is assumed to be axisymmetric and has a density profile corresponding to the theoretical Minimum Mass Solar Nebula \citep[MMSN, see][]{haya81}. Planetesimals have sizes $1\leq s \leq 10\,$km \citep[modified from][]{theb11}.
}
\label{gasnom}
\end{figure*}

\begin{figure}
\includegraphics[width=0.9\columnwidth]{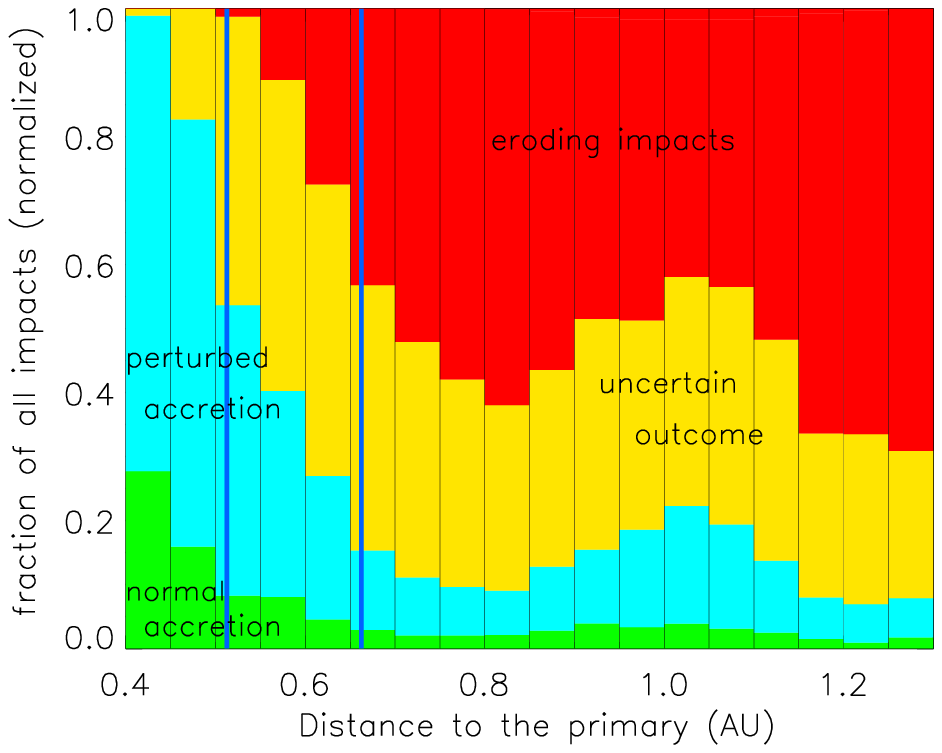}
\caption[]{Planetesimal disc around $\alpha$ Cen B. Numerical simulations with gas drag. Relative importance of different type of mutual impacts as a function of radial distance. \emph{Red}: impacts for which $\Delta v_{s_1,s_2}\geq v_{ero-M}$, where $v_{ero-M}$ is the threshold velocity beyond which an impact between two objects of sizes $s_1$ and $s_2$ always results in net mass loss. \emph{Yellow}: $v_{esc-m}\leq\Delta v_{s_1,s_2}\leq v_{ero-M}$, where  $v_{esc-m}$ is the erosion threshold considering poorly collision-resistant material.
\emph{Green}: $\Delta v_{s_1,s_2}\leq v_{esc}$ where $v_{esc}$ is the escape velocity of the ($s_1,s_2$) pair. Accretion can here proceed unimpeded, in a "runaway growth" way, as around a single star. \emph{Light blue}: $v_{esc}\leq\Delta v_{s_1,s_2}\leq v_{ero-m}$. Collisions result in net accretion, but $\Delta v$ are high enough to cancel off the fast-runaway growth mode.
The two thick blue lines denote the location of the inner limit of the "empirical" and "narrow" Habitable Zones (see Sec.\ref{habit}). The planetesimal size distribution is assumed to be a Maxwellian centered on 5\,km \citep[modified from][]{theb09}.
}
\label{acenhist}
\end{figure}

The presence of a gas disc radically alters this simple picture. As shown by \citet{marz00}, gas drag progressively damps the secular oscillations and planetesimals converge towards an asymptotic equilibrium orbit $e_g$ and $\omega_g$ \citep{paard08}. However, this gas-induced evolution is \emph{size dependent}, so that planetesimals of different sizes end up on different orbits \citep{theb04,theb06}. 
This is clearly illustrated in Fig.\ref{gasnom}, which also shows how small bodies reach their equilibrium orbits faster than larger objects (in the present example, 1\,km objects are already at $e_g$ and $\omega_g$, while 10\,km ones still undergo residual secular oscillations).
As a consequence, $\Delta v$ are small between equal-sized objects but can reach very high values for bodies of different sizes. For most "reasonable" size distributions within the planetesimal population, the differential phasing effect is expected to be the dominant one, and the dynamical environment can thus be strongly hostile to planetesimal accretion in vast regions of the circumstellar disc \citep{theb06,theb11}. These accretion-hostile regions are in general much more extended than the region of orbital stability.
Furthermore, even in the regions where planetesimal accretion $is$ possible, it can often not proceed in the same way as around a single star, because the $\Delta v$ increase is still enough to slow down and impede the runaway growth mode. A worrying result is that, for the emblematic $\gamma$ Cephei and HD196885 cases, the locations at which the planets are observed are probably too perturbed to allow for this planetesimal accretion stage to proceed \citep{paard08, theb11}. As for the arguably most famous binary system, $\alpha$ Centauri, simulations have shown that planetesimal accretion is very difficult in the habitable zone \citep{theb08,theb09}. See for instance Fig.\ref{acenhist}, showing that in the disc around $\alpha$ Cen B the whole region beyond $\sim 0.5\,$ AU is globally hostile to km-sized planetesimal accretion. Note, however, that the innermost regions, where the planet $\alpha$ Cen Bb has possibly been detected, are almost unaffected by binary perturbations.

\citet{xie09} and \citet{xie10} showed that a small inclination of a few degrees between the circumprimary gas disc and the binary orbital plane can help accretion. This is because planetesimal orbital inclinations are segregated by size, thus favouring low-$\Delta v$ impacts between equal-sized bodies over high-$\Delta v$ impacts between differently-sized objects. This reduction of high-velocity impacts comes, however, at a price, which is that the collision rate strongly decreases, so that accretion can only proceed very slowly.
Moreover, the results of \citet{frag11} seem to indicate that taking into account the effect of the gas disc's gravity could offset this positive effect of orbital inclination, leading to high $\Delta v$ dynamical environments (see below)

\subsection{More sophisticated models: gas is still a problem }

\begin{figure}
\includegraphics[width=0.9\columnwidth]{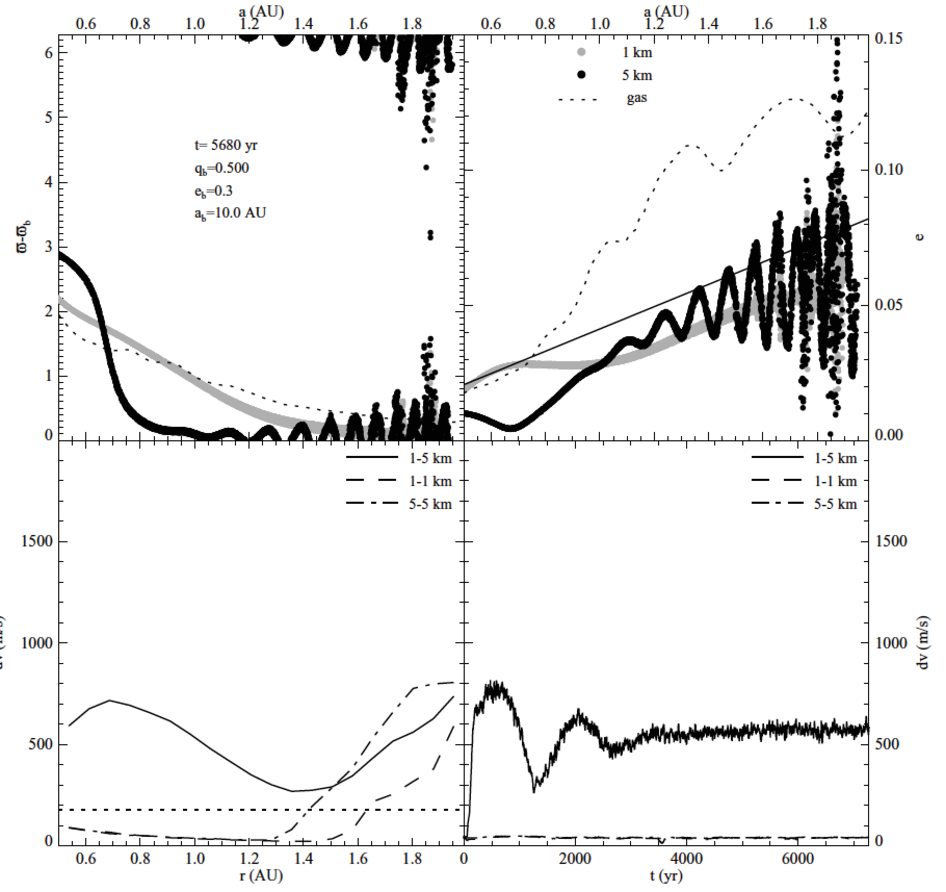}
\caption[]{Planetesimal evolution in an \emph{evolving} gas disc for a tight binary of separation $10\,$AU. Top left: longitude of periastron, after 5680 years, for 1 km planetesimals, 5km ones, and for the gas disc. Top right: orbital eccentricity. Bottom left: encounter velocities between planetesimals as a function of their sizes. Bottom right: average encounter velocities at 1AU as a function of time \citep[from][ courtery of MNRAS]{paard08}
}
\label{paard}
\end{figure}

This globally negative effect of the gas on planetesimal-accretion had been identified in studies considering fiducial static and axisymmetric gas discs, and one could not rule out that this result could be an artifact due to the simplified gas disc prescription. In recent years, new studies have investigated this issue by considering increasingly sophisticated gas-disc models. For the most part, these studies have confirmed the accretion-hostile effect of the gas.
\citet{paard08} have considered a system with an evolving gas disc that also feels the pull of the binary. They have shown that the situation gets even worse for planetesimal accretion: gas streamlines follow paths that are very different from the planetesimal orbits, so that gas drag is enhanced, and so is the accretion-hostile differential phasing effect (see Fig.\ref{paard}).

Another important mechanism that had been neglected in most early planetesimal+gas studies is that of the gas disc's gravity. Due to the difficulty of incorporating this effect in numerical codes, there have only been two attempts at numerically investigating the role of disc self-gravity in circumprimary discs, and only for very limited populations of test planetesimals. The pioneering study by \citet{kley07} has shown that, for a large fraction of the set-ups they explored, the effect of gas-disc gravity can dominate that of gas drag in controlling planetesimal dynamics. A later study by the same team \citep{frag11} found that the net effect of disc gravity is to further increase impact speeds between planetesimals, and this even between equal-sized bodies.
The recent analytical exploration of \citet{rafi13} has nevertheless shown that, if the gas disc is very massive and almost axisymmetric, then its gravity could in fact reduce encounter velocities amongst planetesimals. However, this massive axisymmetric disc prerequisite is probably not likely to be generic, as all hydrodynamical studies of gas discs in binaries have shown that they reach eccentricities that are at least one order of magnitude higher than the few $10^{-3}$ needed in the Rafikov(2013) model \citep{kley07,marz09,marz12,zsom11,mull12}.

On a more positive note, \citet{xie08} showed that, during the late stages when gas dissipates from the disc, the dispersal of planetesimal orbits narrows and the systems can get accretion-friendly again. However, this behaviour probably occurs too late, once most planetesimals have already been either grounded to dust or spiralled onto the star because of gas friction \citep{theb08}.

\section{Late Stages} \label{late}

\begin{figure}
\includegraphics[width=.9\columnwidth]{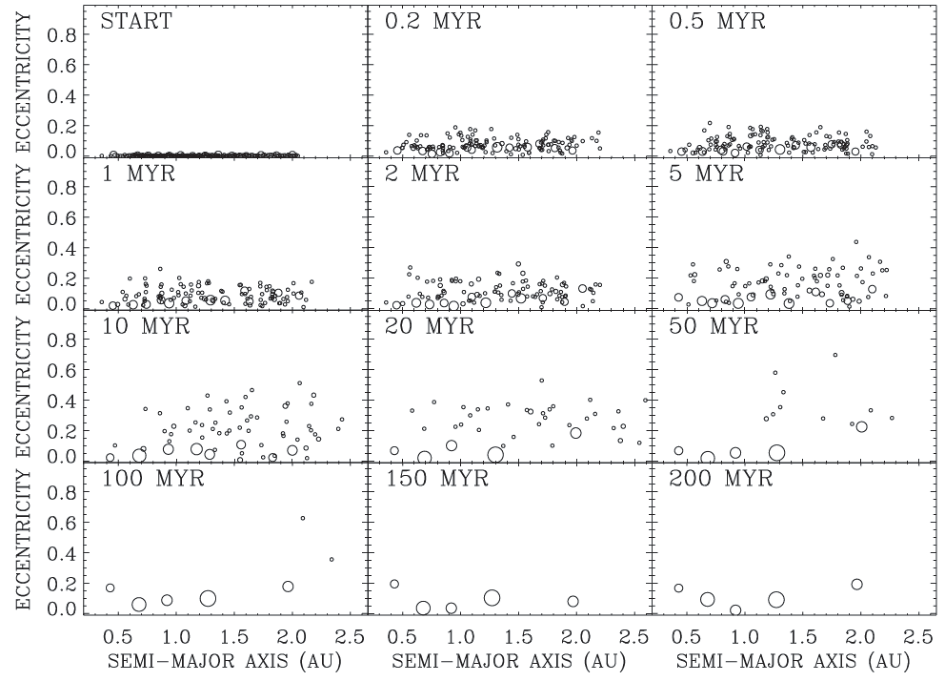}
\caption[]{Late stages of planet formation around the primary of an $\alpha\,$Centauri like binary.
The initial disc consists of a mixture of 14 already formed embryos of Lunar Mass and 140 massive planetesimals of mass $9.33\times 10^{-3}M_{\oplus}$. Results taken from \citet{quin02}, courtesy of the Astrophysical Journal.
}
\label{quin}
\end{figure}

The stage for which the influence of a companion is probably best understood is the final step of planetary accretion, leading from Lunar-sized embryos to fully formed planets. Several studies have shown that the regions where embryo accretion can proceed roughly correspond to those for orbital stability \citep{barb02,quin02,quin07,hagh07,gued08,hagh10}. 
This is a further reassuring result for all known exoplanets-in-binaries. 
As an illustration, we show in Fig.\ref{quin} the results obtained by \citet{quin02} for the late stages of planet formation around a $\alpha$Cen like binary. As can be clearly seen, even in such a tight binary, planetary objects can grow within $\sim 2\,$AU region from the primary star

\begin{figure}
\includegraphics[width=.9\columnwidth]{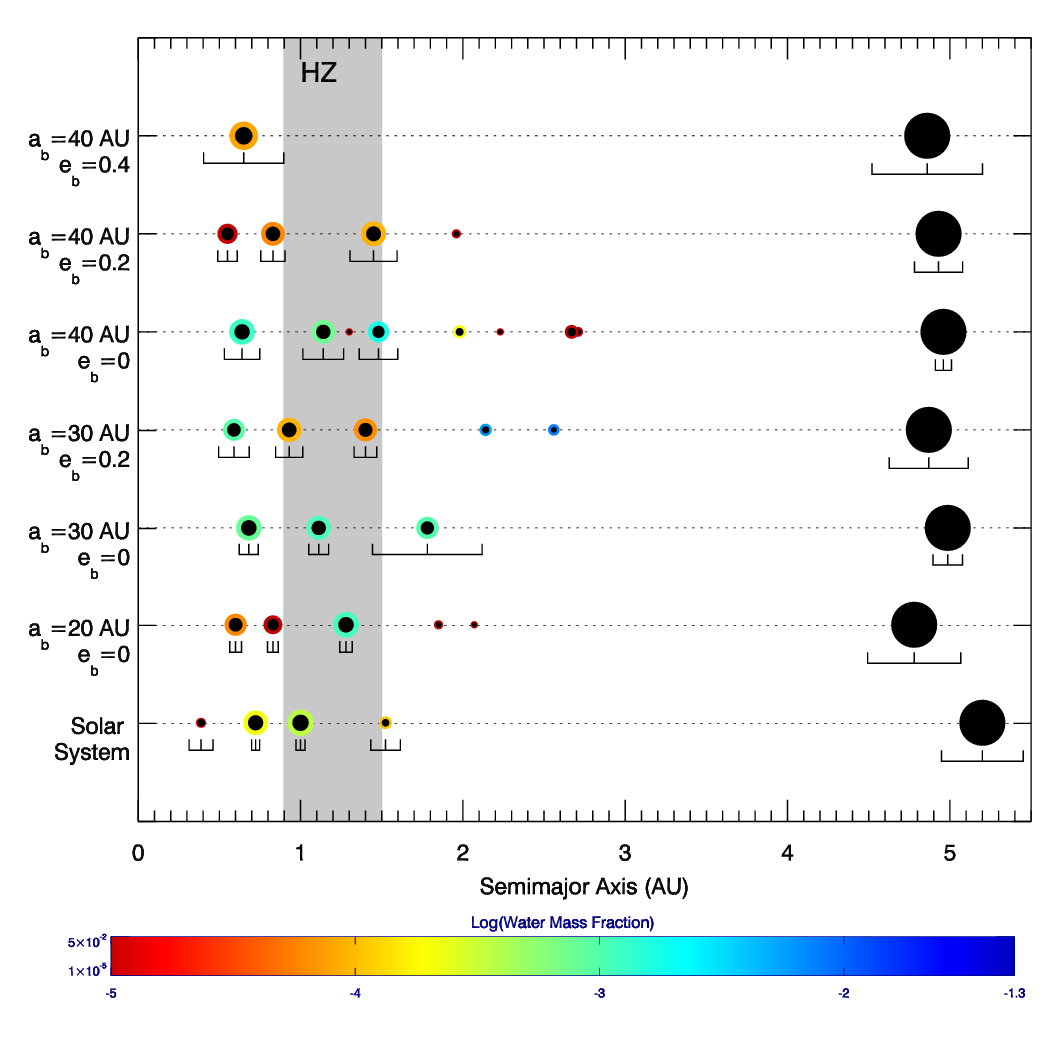}
\caption[]{Late stages of planet formation: Terrestrial planet formation and water delivery in the presence of an already formed giant planet. Results of simulations in a binary system with a mass-ratio of 0.5 and for different values 
of the eccentricity $(e_b)$ and semimajor axis $(a_b)$ of the binary. The solar system's configuration is given as a comparision \citep[taken from][ courtesy of the Astrophysical Journal]{hagh07}.
}
\label{raym1}
\end{figure}

More recent work on these late stages have been devoted to more specific issues. As an illustration, we present here results obtained by \citet{hagh07} that focuses on the level of water delivery and water mixing in the terrestrial regions within a binary, assuming that a giant planet could form further out in the disc. The authors considered a binary with a solar-mass primary and a Jupiter-mass giant planet at 5 AU. The mass of the secondary star was taken to be 0.5, 1.0, and 1.5 solar-masses and the semimajor axis and eccentricity of the binary were varied in the range of  20-40 AU and 0-0.4, respectively. For each value of the mass of the primary, Haghighipour \& Raymond (2007) used the planetary orbit stability criterion given by Rabl \& Dvorak (1988) and Holman \& Wiegert (1999), and identified the combination of the mass, semimajor axis, and eccentricity of the binary for which the giant planet would maintain a stable orbit at 5 AU. They then considered a disc of 115 Moon-to-Mars-sized bodies with masses ranging from 0.01 to 0.1 Earth masses.

\begin{figure}
\includegraphics[width=.9\columnwidth]{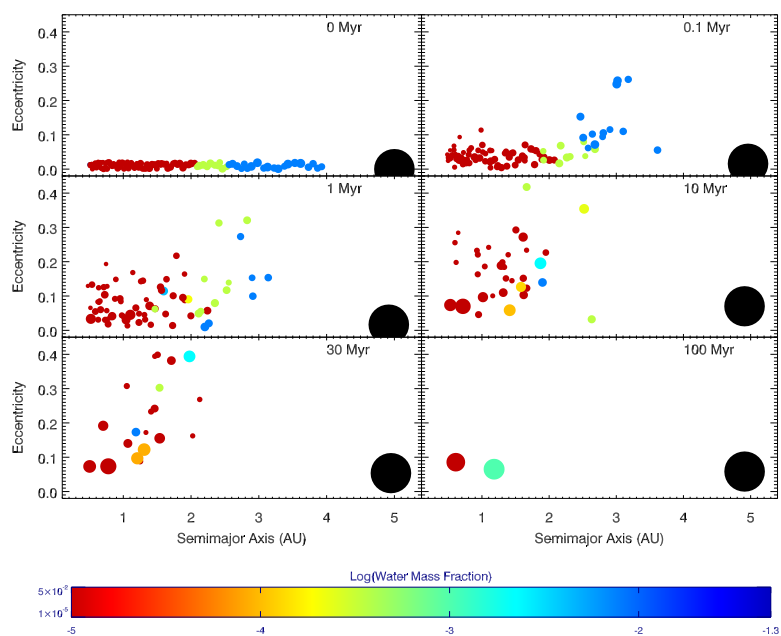}
\vskip 30pt
\caption[]{Temporal evolution of the terrestrial planet growth in a binary system with a mass-ratio of 0.5 and for different values of the eccentricity $(e_b)$ and semimajor axis $(a_b)$ of the binary \citep[taken from][ courtesy of the Astrophysical Journal]{hagh07}.
}
\label{raym2}
\end{figure}

Fig.\ref{raym1} shows the results of a sample of their simulations for a binary with a mass-ratio of 0.5 and for different values of its semimajor axis and eccentricity. The gray area corresponds to the boundaries of the habitable zone of the primary star (Kasting et al. 1993). As a point of comparison, the inner planets of the solar system are also shown. As shown here, many planets of the same or similar size as Earth and with substantial amount of water formed in and around 1 AU. Fig.\ref{raym2} shows the snapshots of one of such simulations. During the course of the integration, the embryos in part of the disk close to the giant planet are dynamically excited. As a result of the interaction of embryos with one another, the orbital excitation of these objects  is transmitted to other bodies in closer orbits causing many of them to be scattered out of the system or undergo radial mixing. In the simulation of Fig.\ref{raym2}, this process results in formation of a planet slightly larger than Earth and with a large reservoir of water. 

An interesting result of the \citet{hagh07} numerical exploration is that, \emph{assuming the first stages of planet formation were successful}, binaries with moderate to large perihelia and with giant planets on low-eccentricity orbits are most favorable for Earth-like planet formation. Similar to the formation of terrestrial planets around single stars, where giant planets, in general, play destructive roles, a strong interaction between the secondary star and the giant planet in a 
binary-planetary system (i.e., a small binary perihelion) increases the orbital eccentricity of this object, and results in the removal of the terrestrial planet-forming materials from the system.

\section{Planets in formation-hostile regions: solving the paradox} \label{altern}

\begin{figure}
\includegraphics[width=.8\columnwidth]{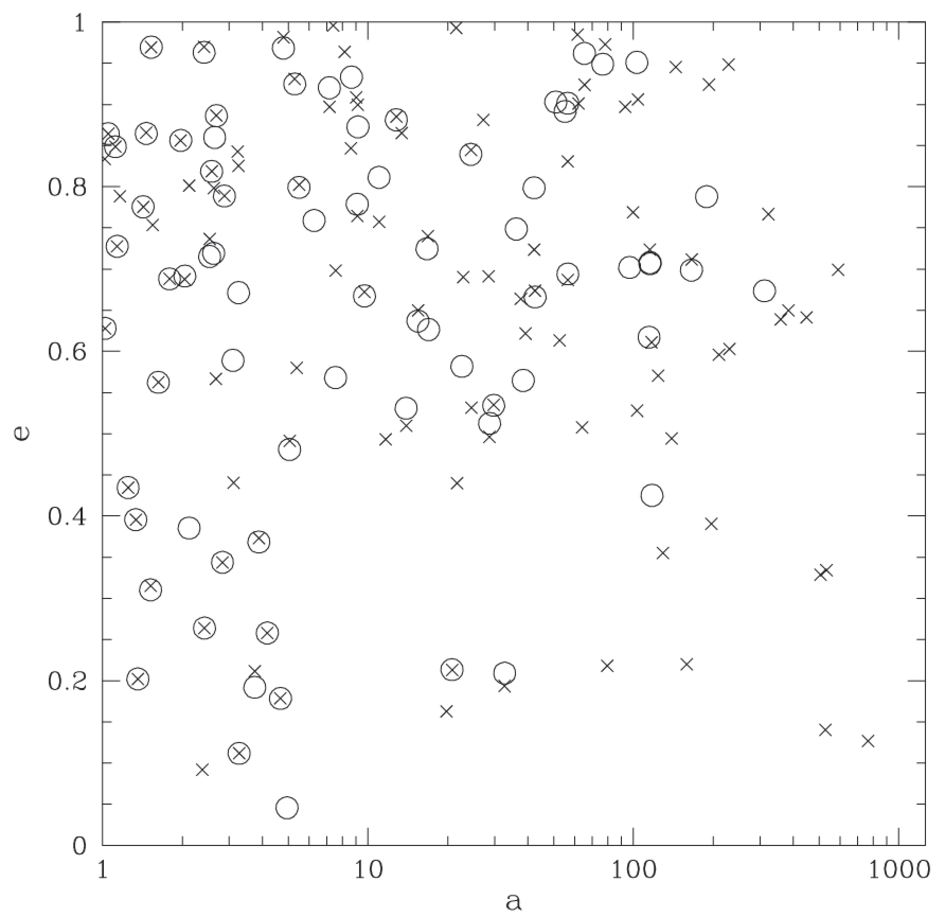}
\caption[]{Initial (crosses) and final (circles) binary orbital parameters obtained in the stellar cluster evolution simulations of \citet{malm07}. Figure reproduced from the Figure 4 of that paper in the MNRAS.}
\label{malm}
\end{figure}

\begin{figure}
\includegraphics[width=.9\columnwidth]{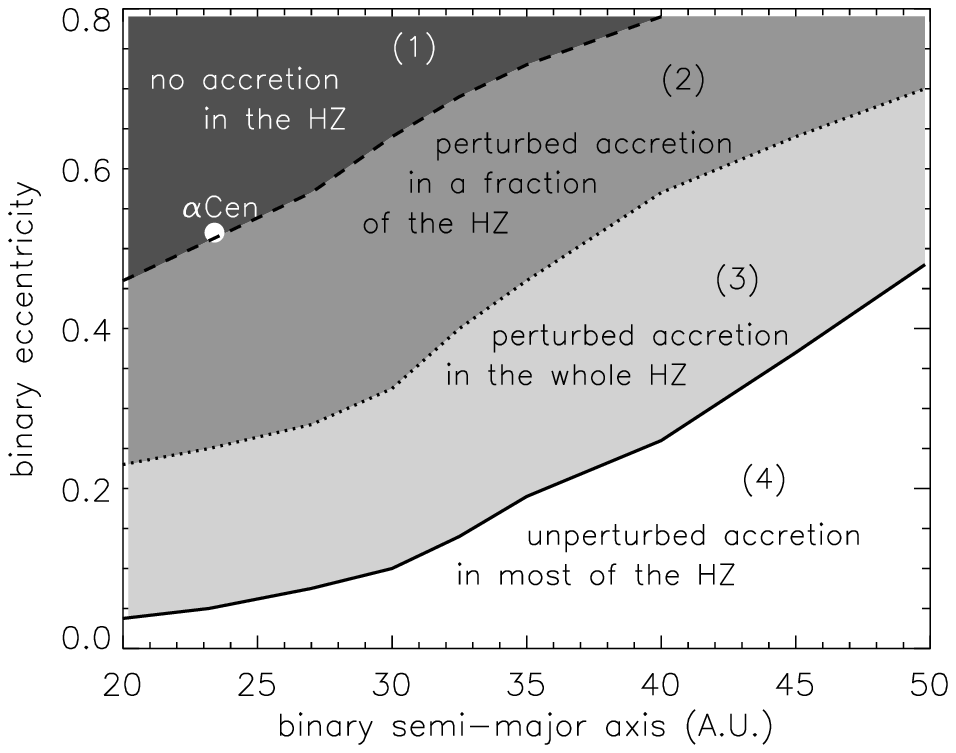}
\caption[]{Accretion vs. Erosion behaviour of a population of kilometre-sized planetesimals in the habitable zone of the $\alpha$ Centauri when varying the binary's separation and excentricity \citep[taken from][ courtesy of the MNRAS]{theb09}.}
\label{orbvaralpha}
\end{figure}

As has been shown in the previous sections, with the exception of its very late stages, the formation of S-type planets in close binaries has many obstacles to overcome. This is in particular true for the planetesimal-accretion stage, which seems unlikely to proceed at the location where the $\gamma$ Ceph, HD196885 and HD41004 planets have been detected. To explain this paradox of having planets in planet-formation hostile environments, several solutions can be considered.

A first possibility is that these planets were not formed in-situ, but in friendlier regions closer to the primary and later migrated outward to their present position. Classical type I or type II migrations in the gas disc would usually move the planet in the "wrong" direction, i.e., inward. There exists, however, possibilities for these migrations to revert direction \citep[e.g.,][]{mass01,pier12}, but these scenarios have not been explored yet in the context of binaries. The outward migration process that has been explored for binaries regards a later phase, i.e., the planetesimal-driven migration of terrestrial embryos once the gas disc has dissipated. \citet{payn09} have shown that such a mechanism can indeed move some large embryos beyond their initial position, over timescales of typically $10^{6}$ years. However, the maximum outward displacement does not exceed $\sim30$\% of the initial sem-major axis, and is not enough to explain the formation of extreme planets such as HD196885 or $\gamma$ Ceph.
Another potential way of moving planets outward is by planet-planet scattering. \citet{marz05} have studied the dynamical evolution of unstable multi-planet systems in binaries and showed that, for 3 neighbouring Jovian planets, a possible outcome is the ejection of one planet, the inward jump of a second one, and the outward displacement of the third one. In this case, one would end up with 2 giant planets, one of which potentially far outside the region where it accreted. However, the major caveat of this scenario is that it requires the presence of a second giant planet on a close circumprimary orbit, and that such a planet should in principle be more easily detected that the outer one. As such, the non-detection of any second Jovian planet in HD196885 or $\gamma$ Cephei is a problematic issue. A related, albeit theoretical problem is that this scenario requires the truncated protoplanetary disc to be massive enough to form not just one but $two$ giant planets.

Another solution is simply to bypass the critical km-sized planetesimals range. If indeed the "initial" planetesimals are large, typically $\geq 100$\,km, then they are big enough to sustain high velocity collisions and can survive and grow in the perturbed environment of a binary. Of course, for this scenario to work, such large planetesimals should be formed directly from much smaller boulder and pebble-sized bodies, so that the whole dangerous 100m-to-10km range is avoided. Interestingly, two recent planetesimal-forming models, the streaming-instability model of \citet{joha07} and the clumping mechanism of \citet{cuzz08}, advocate such a km-size-straddling formation mode. However, it remains to be seen if these formation mechanisms can operate in the highly-perturbed environment of a binary.

Even if the kilometre-sized planetesimal phase cannot be bypassed, recent studies have shown that there is a possibility that these bodies might after all grow even in the presence of high-$\Delta v$ collisions. \citet{xie10b} have analytically investigated the "snowball" growth mode, in which planetesimals accrete mass preferentially by sweeping up of dust particles instead of mutual collisions with other planetesimals. If the mass density contained in dust exceeds that contained in large bodies, then this mechanism could provide a viable growth mode in binaries, because dust-on-planetesimal impacts can result in accretion for $\Delta v$ much higher than planetesimal-on-planetesimal ones. These results have been strengthened by \citet{paard10}, who incorporated for the first time a collisional-evolution model in their simulations, taking into account the fragments produced by high-velocity impacts. Their runs have shown that many planetesimals are able to re-accrete, by sweeping, a large fraction of the mass that has been lost to small fragments by earlier high-$\Delta v$ collisions. Moreover, frequent collisions with these small fragments can prevent km-sized planetesimals from reaching their equilibrium secular+gas-drag imposed orbits, so that their \emph{mutual} impact velocities never fully reach the high values resulting from the differential phasing effect. However, the collision outcome prescription of \citet{paard10} is still relatively simplistic, with all bodies smaller than $\sim 1\,$km being treated at dust coupled to the gas, and more sophisticated models have to been tested.

Another promising explanation for the presence of S-type planets in very close binaries is that these binaries were initially wider than today. As it happens, there is in fact theoretical support for this hypothesis. If indeed most stars are born in clusters, then they should experience many close encounters in their early history. \citet{malm07} have shown that, for a typical stellar cluster, some very wide binaries get broken by close encounters, but for binaries that survive, the main effect of these encounters is to shrink their orbits. Interestingly, for the cluster set-up they considered, \citet{malm07} found that around half of the binaries having a present separation of $\sim 20\,$AU have had a significant orbit-shrinking encounter (see Fig.\ref{malm} \footnote{The simulations of \citet{malm07} were, however, only assuming a single, hypothetical, initial set-up. More generic numerical investigations, exploring a wider range of possible initial conditions, should be carrier out}. However, for the specific case of HD196885, we run a series of simulations, each time increasing the binary separation, and found that the present-day planet location became accretion-friendly only for a binary separation of $\sim45$\,AU. This means that this separation should have shrunk by at least a factor 2 during the system's evolution, and it remains to be seen if such a change is realistic.
Perspectives are however more optimistic for the HZ of the $\alpha$ Centauri system, for which only a moderately larger and/or less eccentric initial binary orbit would be enough to allow planetary accretion there (see Fig.\ref{orbvaralpha}).

If none of the aforementioned solution works, then a more radical explanation should be considered, i.e., that planets in close binaries form by a different channel than the usual core-accretion scenario. An obvious alternative would be the gravitational instability scenario \citep[e.g.][]{boss97,maye10}, in which planets form very quickly by direct disc fragmentation. Although this model still has several issues that need to be solved, in particular the cooling of collapsing proto-planetary clumps, it has recently been advocated as a possible explanation for the presence of giant planets at large radial distances from their stars, such as in the HR8799 system \citep{boss11}. 
\citet{duch10} argues that this alternative formation mechanism could be at play in $a_b\leq 100\,$AU binaries, and that it could explain why planets in tight binaries are more massive than around single stars. In such close binaries, the instability mechanism could indeed be more efficient than around single stars, because circumprimary discs might be more compact, and thus more prone to fragmentation, than non-truncated ones, but also because binary perturbations could act as an additional trigger to instabilities. As a result, even close-in planets could be formed this way in binaries \citep{duch10}. 
However, several studies have also shown that the instability scenario does also encounter major difficulties in the context of close binaries, and that no circumprimary disc gets dense enough to be unstable. See, for example, \citet{nels00} or \citet{maye05, maye10}, whose main conclusions are that the presence of a close ($\leq$ 50--60\,AU) companion could greatly hinder the development of instabilities. This issue is thus far from being settled yet, and further, more detailed investigations are needed to assess if disc-fragmentation can be considered as a viable alternative formation scenario in close binaries.


\section{Habitability} \label{habit}

Although Earth-like planets are yet to be discovered in the habitable zone (HZ) of binary star systems, many planet-formation-in-binaries studies have rightfully payed special attention to the HZ (see previous sections).
In most of these simulations, it has been generally assumed that the HZ of a binary is equivalent to the single-star HZ of its planet-hosting star. Although in binaries with separations smaller than 50 AU, the secondary star plays an important role in the formation, long-term stability, and water content of a planet in the HZ of the primary, 
the effect of the secondary on the range and extent of the HZ in these systems was ignored. However, the fact that this star can affect planet formation around the primary, and can also perturb the orbit of a planet in the primary's HZ in binaries with moderate eccentricities, implies that the secondary may play a non-negligible role in the habitability of 
the system as well.

Unlike around single stars where the HZ is a spherical shell with a distance determined by the host star alone, in binary star systems, the radiation from the stellar companion can influence the extent and location of the HZ of the system. Especially for planet-hosting binaries with small stellar separations and/or in binaries where the planet orbits the less luminous star, the amount of the flux received by the planet from the secondary star may become non-negligible. 

In addition, effects such as the gravitational perturbation of the secondary star (see e.g., Georgakarakos 2002, Eggl et al. 2012) can influence a planet's orbit in the Binary HZ and lead to temperature fluctuations if the planetary atmosphere cannot buffer the change in the combined insolation. Since in an S-type system, the secondary orbits more slowly than the planet, during one period of the binary, the planet may experience the effects of the secondary several times. The latter, when combined with the atmospheric response of the planet, defines the HZ of the system. 

Despite the fact that as a result of the orbital architecture and dynamics of the binary, at times the total radiation received by the planet exceeds the radiation that it receives from its parent star alone by a non-negligible amount, the boundaries of the actual HZ of the binary cannot be obtained by a simple extrapolation of the boundaries of the HZ of its planet-hosting star. Similar to the HZ around single stars, converting from insolation to equilibrium temperature of the planet depends strongly 
on the planet's atmospheric composition, cloud fraction, and star's spectral type. A planet's atmosphere responds differently to stars with different spectral distribution of incident energy. Different stellar types will therefore contribute differently to the total amount of energy absorbed by the planetary atmosphere (see e.g., Kasting et al. 1993). A complete and realistic 
calculation of the HZ has to take into account the spectral energy distribution (SED) of the binary stars as well as the planet's atmospheric response. In this section, we address these issues and present a coherent and self-consistent model for determining the boundaries of the HZ of S-type binary systems.

\subsection{Calculation of the Binary Habitable Zone}

Habitability and the location of the HZ depend on the stellar flux at the planet's location as well as the planet's atmospheric composition. The latter determines the albedo and the greenhouse effect in the planet's atmosphere and as such plays a strong role in determining the boundaries of the HZ. Examples of atmospheres with different chemical compositions include the original CO$_2$/H$_2$O/N$_2$ model (Kasting et al 1993; Selsis et al. 2007; Kopparapu et al. 2013a) with a water reservoir like 
Earth's, and model atmospheres with high H$_2$/He concentrations (Pierrehumbert \& Gaidos 2011) or with limited water supply (Abe et al. 2011).
 
At present, the recent update to the Sun's HZ given by Kopparapu et al. (2013a\&b) presents the best model. According to this model, the HZ is an annulus around a star where a rocky planet with a CO$_2$/H$_2$O/N$_2$ atmosphere and sufficiently large water content (such as on Earth) can host liquid water permanently on its solid surface. This definition of the HZ assumes the abundance of CO$_2$ and H$_2$O in the atmosphere is regulated by a geophysical cycle similar to Earth's carbonate silicate cycle. The inner and outer boundaries of the HZ in this model are associated with a H$_2$O-- and CO$_2$--dominated atmosphere, respectively. Between those limits on a geologically active planet, climate stability is provided by a feedback mechanism in which atmospheric CO$_2$ concentration varies inversely with planetary surface temperature. 

The locations of the inner and outer boundaries of a single star's as well as a binary's HZ depend also on the cloud fraction in the planet's atmosphere. That is because the overall planetary albedo is a function of the chemical composition of the clear atmosphere as well as the additional cooling or warming of the atmosphere due to clouds. Since the model by Kopparapu et al. (2013a\&b) does not include clouds, it is customary to define two types of HZ; the {\it narrow} HZ which is considered to be the region between the limits of runaway and maximum greenhouse effect in the model by Kopparapu et al (2013a\&b), and the {\it empirical} HZ, as a proxy to the effect of clouds, that is derived using the fluxes received by Mars and Venus at 3.5 and 1.0 Gyr, respectively (the region between Recent Mars and Early Venus). 
At these times, the two planets do not show indications for liquid water on their surfaces (see Kasting et al. 1993). In these definitions, the locations of the HZs are determined based on the flux received by the planet (see e.g., Kasting et al. 1993; Selsis et al. 2007; Kaltenegger \& Sasselov 2011; and Kopparapu et al. 2013a).

\subsection{Effect of Star's Spectral Energy Distribution (SED)}

The locations of the boundaries of the HZ depend on the flux of the star at the orbit of the planet. In a binary star system where the planet is subject to radiation from two stars, the flux of the secondary star has to be added to that of the primary (planet-hosting star) and the total flux can then be used to calculate the boundaries of the HZ. However, because the response of a planet's atmosphere to the radiation from a star depends strongly on the star's SED, a simple summation of fluxes is not applicable. The absorbed fraction of the absolute incident flux of each star at the top of the planet's atmosphere will differ 
for different SEDs. Therefor in order to add the absorbed flux of two different stars and derive the limits of the HZ for a binary system, one has to weight the flux of each star according to the star's SED. The relevant flux received by a planet in this case is the sum of the  spectrally weighted stellar flux, separately received from each star of the binary, as given by [see Kaltenegger \& Haghighipour (2013) for details],

\begin{equation}
{F_{\rm Pl}}(f,{T_{\Pr}},{T_{\rm Sec}})\,=\,
{W_{\rm Pr}}(f,{T_{\rm Pr}})\, {{{L_{\rm Pr}}({T_{\rm Pr}})}\over {r_{\rm Pl-Pr}^2}}\,+\,
{W_{\rm Sec}}(f,{T_{\rm Sec}})\, {{{L_{\rm Sec}}({T_{\rm Sec}})}\over {r_{\rm Pl-Sec}^2}}\,.
\end{equation}

\noindent
In this equation, $F_{\rm Pl}$ is the total flux received by the planet, $L_i$ and $T_i$ ($i$=Pr, Sec) represent the luminosity and effective temperature of the primary and secondary stars, $f$ is the cloud fraction of the planet's atmosphere, and ${W_i}(f,{T_i})$ is the binary stars' spectral weight factor. The quantities $r_{\rm Pl-Pr}$ and $r_{\rm Pl-Sec}$ in equation (1) represent the distances between the planet and the primary and secondary stars, respectively (figure 11). In using equation (1), we normalize the weighting factor to the flux of the Sun.

\begin{figure*}
\makebox[\textwidth]{
\includegraphics[width=.6\columnwidth]{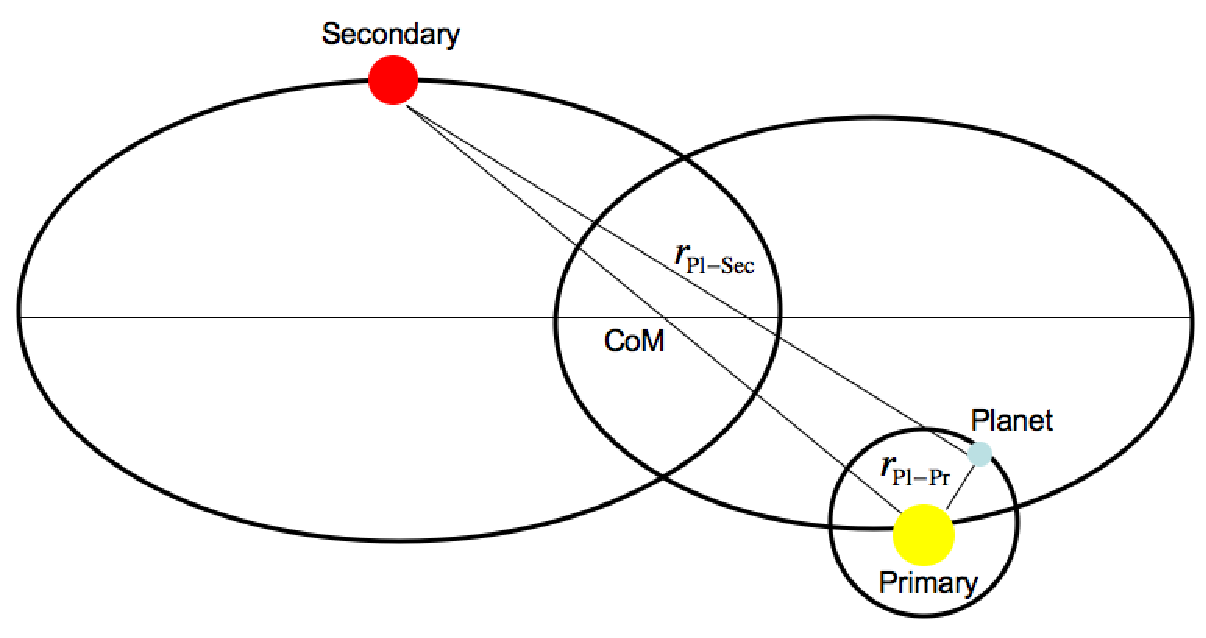}
\hfil
\includegraphics[width=.45\columnwidth]{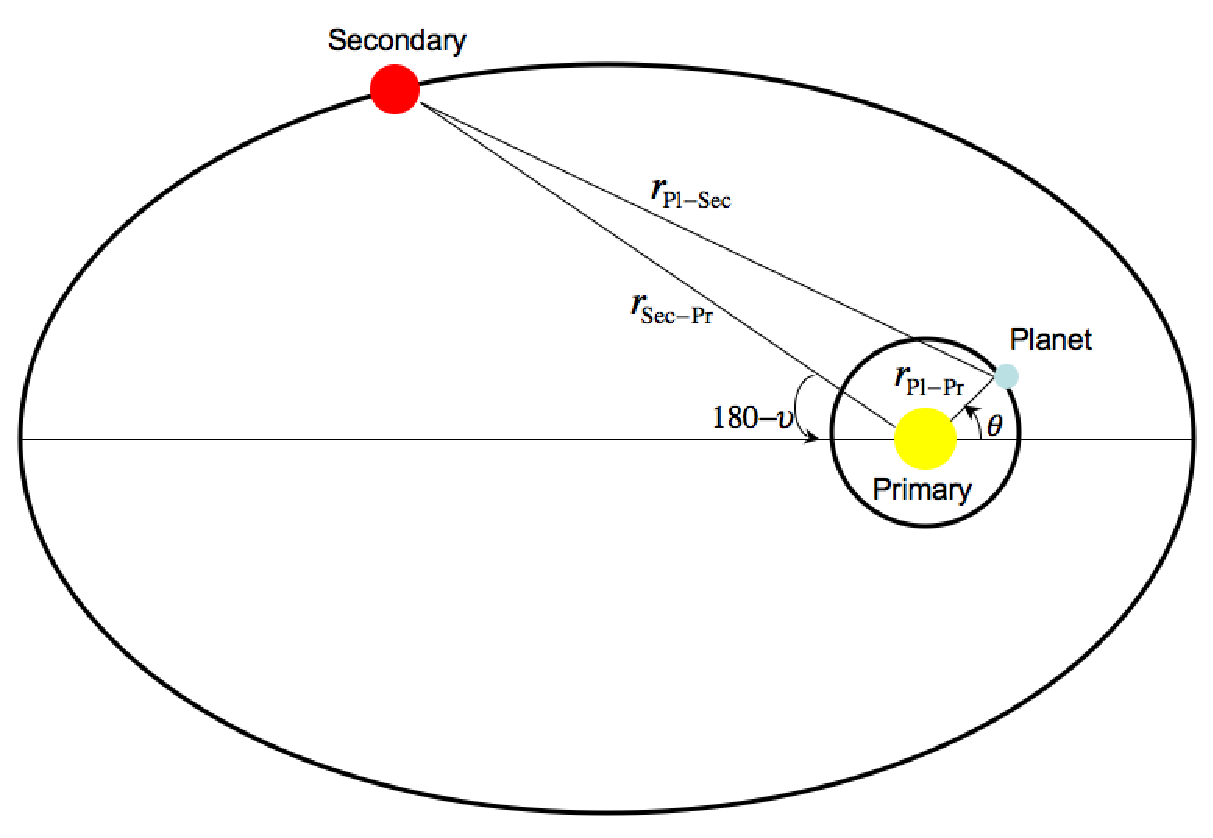}
}
\caption[]{\emph{Left panel}: Schematic presentation of an S-type system. The two stars of the binary, primary and secondary, revolve around their center of mass (CoM) while the planet orbits only one of the stars (top panel). It is, however, customary to
neglect the motion of the binary around its CoM and consider the motion of the secondary around a stationary primary (bottom panel).
}
\label{schema}
\end{figure*}

From equation (1), the boundaries of the HZ of the binary can be defined as distances where the total flux received by the planet is equal to the flux that Earth receives from the Sun at the inner and outer edges of its HZ. Since in an S-type system, the planet revolves around one star of the binary, we determine the inner and outer edges of the HZ with respect to the planet-hosting star. As shown in figure 11, it is customary to consider the primary of the system to be stationary, and calculate the orbital elements with respect to the stationary primary star. 
In the rest of this section, we will follow this convention and consider the planet-hosting star to be the primary star as well. In that case, the range of the HZ of the binary can be obtained from 

\begin{equation}
{W_{\rm Pr}}(f,{T_{\rm Pr}})\, {{{L_{\rm Pr}}({T_{\rm Pr}})}\over {l_{\rm x-Bin}^2}}\,+\,
{W_{\rm Sec}}(f,{T_{\rm Sec}})\, {{{L_{\rm Sec}}({T_{\rm Sec}})}\over {r_{\rm Pl-Sec}^2}}\,=\,
{{L_{\rm Sun}}\over {l_{\rm x-Sun}^2}}\,.
\end{equation}

\noindent
In equation (2), the quantity ${l_{\rm x}}$ represents the inner and outer edges of the HZ with x=(in,out). As mentioned earlier, the values of $l_{\rm in}$ and $l_{\rm out}$ are model-dependent and change for different values of cloud fraction, $f$, and atmosphere composition.

\subsection{Calculation of Spectral Weight Factors}

To calculate the spectral weight factor $W(f,T)$ for each star of the binary and in terms of their SEDs, we calculate the stellar flux at the top of the atmosphere of an Earth-like planet at the limits of the HZ, in terms of the stellar effective temperature. To determine the locations of the inner and outer boundaries of the HZ of a main sequence star with an effective temperature of $2600 K < {T_{\rm Star}} < 7200$ K, we use equation (3) (see Kopparapu et al 2013a)

\begin{equation}
{l_{\rm x-Star}}\,=\,{l_{\rm x-Sun}}\, 
\Biggl[{{L/{L_{\rm Sun}}}\over{1+{\alpha_{\rm x}} ({T_i})\,{l_{\rm x-Sun}^2}}} \Biggr]^{1/2}
\end{equation}

\noindent
In this equation, ${l_{\rm x}}=({l_{\rm in}},{l_{\rm out}})$ is in AU, ${T_i}{\rm (K)}={T_{\rm Star}}{\rm (K)}-5780$, and 

\begin{equation}
{\alpha_{\rm x}}({T_i})\,=\,{a_{\rm x}}{T_i}\,+\,{b_{\rm x}}{T_i^2}\,+\,{c_{\rm x}}{T_i^3}\,+\,{d_{\rm x}}{T_i^4}\,,
\end{equation}

\noindent
where the values of coefficients ${a_{\rm x}}, {b_{\rm x}}, {c_{\rm x}}$, ${d_{\rm x}}$, and $l_{\rm x-Sun}$ are given in Table 1 (see Kopparapu et al. 2013b). From equation (3), the flux received by the planet from its host star at the limits of the Habitable Zone can be calculated using equation (5). The results are given in Table 1;

\begin{equation}
{F_{\rm x-Star}}\big(f, {T_{\rm Star}}\big)\,=\,{F_{\rm x-Sun}}(f,{T_{\rm Star}})\,
\bigg[1\,+\,{\alpha_{\rm x}}({T_i})\,{l_{\rm x-Sun}^2}\bigg]\,.
\end{equation}

\noindent
From equation (5), the spectral weight factor $W(f,T)$ can be written as

\begin{equation}
{W_i}(f,{T_i})\,=\,\biggl[1\,+\,{\alpha_{\rm x}}({T_i})\, {l_{\rm x-Sun}^2}\,\biggr]^{-1}\,.
\end{equation}

\noindent
Table 2 and figure 12 show $W(f,T)$ as a  function of the effective temperature of a main sequence planet-hosting star for the narrow (left panel) and empirical (right panel) boundaries of the HZ. As expected, hotter stars have weighting factors smaller than 1 whereas the weighting factors of cooler stars are larger than 1.

\begin{table}
\centering
\caption{Values of the coefficients of equation (4). See Kopparapu et al. (2013b) for details.}
\begin{tabular}{lcccc}
       &  \multicolumn{2}{c}{Narrow HZ}  & \multicolumn{2}{c}{Empirical HZ}  \\
\hline
      & Runaway Greenhouse &  Maximum Greenhouse  & Recent Venus & Early Mars \\   
\hline
$l_{\rm {x-Sun}}$ (AU) & 0.97  & 1.67  &  0.75  &   1.77   \\
Flux (Solar Flux $@$ Earth) & 1.06  & 0.36  & 1.78 & 0.32 \\
a    & $1.2456 \times {10^{-4}}$   & $5.9578 \times {10^{-5}}$   & $1.4335 \times {10^{-4}}$   & $5.4471 \times {10^{-5}}$ \\
b    & $1.4612 \times {10^{-8}}$   & $1.6707 \times {10^{-9}}$   & $3.3954 \times {10^{-9}}$   & $1.5275 \times {10^{-9}}$ \\
c    & $-7.6345 \times {10^{-12}}$ & $-3.0058 \times {10^{-12}}$ & $-7.6364 \times {10^{-12}}$ & $-2.1709 \times {10^{-12}}$ \\
d    & $-1.7511 \times {10^{-15}}$ & $-5.1925 \times {10^{-16}}$ & $-1.1950 \times {10^{-15}}$ & $-3.8282 \times {10^{-16}}$ \\
\end{tabular}
\end{table}

\begin{figure*}
\makebox[\textwidth]{
\includegraphics[width=.5\columnwidth]{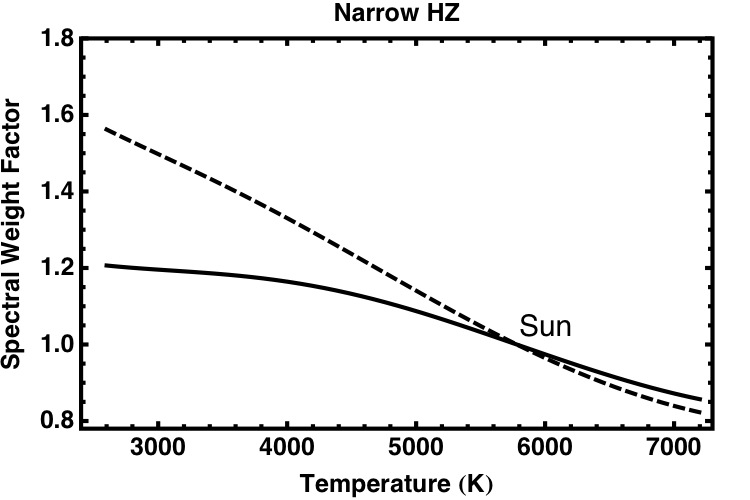}
\hfil
\includegraphics[width=.5\columnwidth]{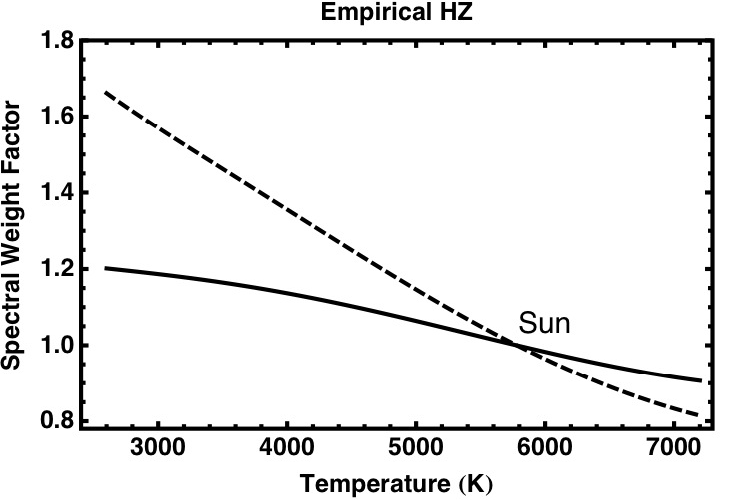}
}
\caption[]{Spectral weight factor $W(f,T)$ as a function of stellar effective
temperature for the narrow (left) and empirical (right) HZs. The solid
line corresponds to the inner and the dashed line corresponds to the outer
boundaries of HZ.We have normalized $W(f,T)$ to its solar value, indicated on the
graphs (Sun).
}
\label{SWF}
\end{figure*}

\subsection{Effect of Binary Eccentricity}

To use equation (2) to calculate the boundaries of the HZ, we assume that the orbit of the (fictitious) Earth-like planet around its host star is circular. In a close binary system, the gravitational effect of the secondary may deviate the motion of the planet from a circle and cause its orbit to become eccentric. In a binary with a given semimajor axis and mass-ratio, the eccentricity has to stay within a small to moderate level to avoid strong interactions between the secondary star and the planet, and to allow the planet to maintain a long-term stable orbit (with a low eccentricity) in the primary's HZ. The binary eccentricity itself is constrained by the fact that in highly eccentric systems, periodic close approaches of the two stars truncate their circumstellar disks depleting them from planet-forming material (Artymowicz \& Lubow 1994) and restricting the delivery of water-carrying objects to an accreting terrestrial planets in the Binary HZ (Haghighipour \& Raymond 2007).

This all implies that in order for the binary to be able to form a terrestrial planet in its HZ, its eccentricity cannot have large values. In a binary with a small eccentricity, the deviation of the planet's orbit from circular is also small and appears in the form of secular changes with long periods (see e.g. Eggl et al. 2012). Therefore, to use equation (2), one can approximate the orbit of the planet by a circle without the loss of generality.
      
The habitability of a planet in a binary system also requires long-term stability in the HZ. For a given semimajor axis ${a_{\rm Bin}}$, eccentricity ${e_{\rm Bin}}$, and mass-ratio $\mu$ of the binary, there is an upper limit for the semimajor axis of the planet beyond which the perturbing effect of the secondary star will make the orbit of the planet unstable. This maximum or {\it critical} semimajor axis $({a_{\rm Max}})$ is given by (Rabl \& Dvorak 1988, Holman \& Wiegert 1999)

\begin{equation}
{a_{\rm Max}}\,=\,{a_{\rm Bin}}\Big(0.464 - 0.38\, \mu - 0.631\, {e_{\rm Bin}} + 
0.586 \, \mu \, {e_{\rm Bin}} + 0.15 \, {e_{\rm Bin}^2} - 0.198\, \mu \,{e_{\rm Bin}^2}\Big)\,.
\end{equation} 

\noindent
In equation (7), $\mu = {m_2}/({{m_1}+{m_2}})$ where ${m_1}$ and $m_2$ are the masses of the primary (planet-hosting) and secondary stars, respectively. One can use equation (7) to determine the maximum binary eccentricity that would allow the planet to have a stable orbit in the HZ (${l_{\rm out}} \leq {a_{\rm Max}}$). For any smaller value of the binary eccentricity, the entire HZ will be stable.

\section{Examples of the Habitable Zone of Main Sequence S-Type Binaries}

As mention earlier, we assume that the orbit of the planet around its host star is circular. Without knowing the exact orbital configuration of the planet, one can only estimate the boundaries of the Binary HZ by calculating the maximum and minimum additional flux from the secondary star at its closest and furthest distances from a fictitious Earth-like planet, as a first order approximation. Note that using the maximum flux of the secondary onto the planet for calculating the new Binary HZ overestimates the shift of the HZ from the single star's HZ to the Binary HZ due to the secondary because the planet's atmosphere can buffer an increase in radiation temporarily. This shift is underestimated when one uses the minimum flux received from the secondary star onto the planet. To improve on this estimation, one needs to know the orbital positions of the planet as well as the stars in the binary. That way one can determine the exact flux over time reaching the planet as well as the number of planetary orbits over which the secondary's flux can be averaged. This depends on the system's geometry (both stellar and planetary parameters) and needs to be calculated for each planet hosting S-type system, individually.

\begin{figure*}
\makebox[\textwidth]{
\includegraphics[width=.5\columnwidth]{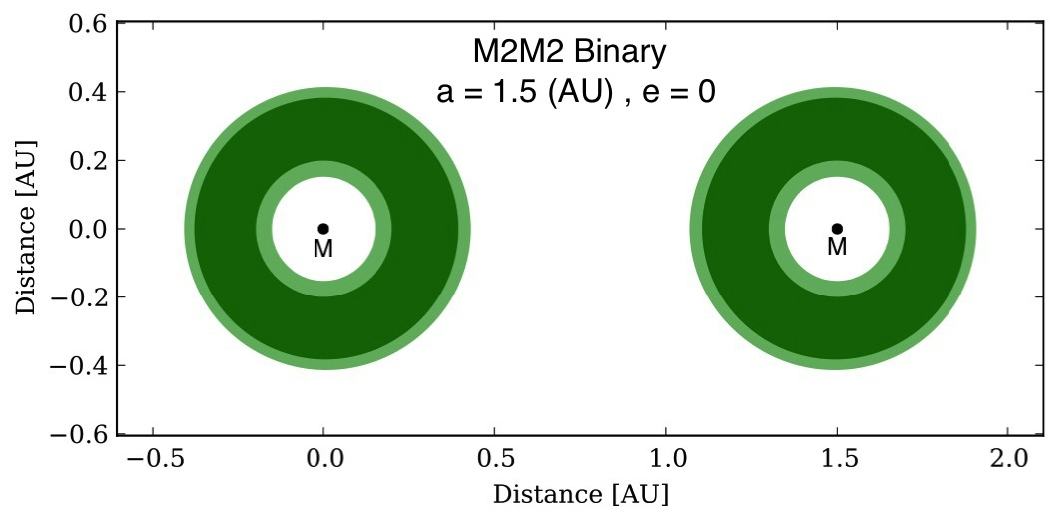}
\hfil
\includegraphics[width=.5\columnwidth]{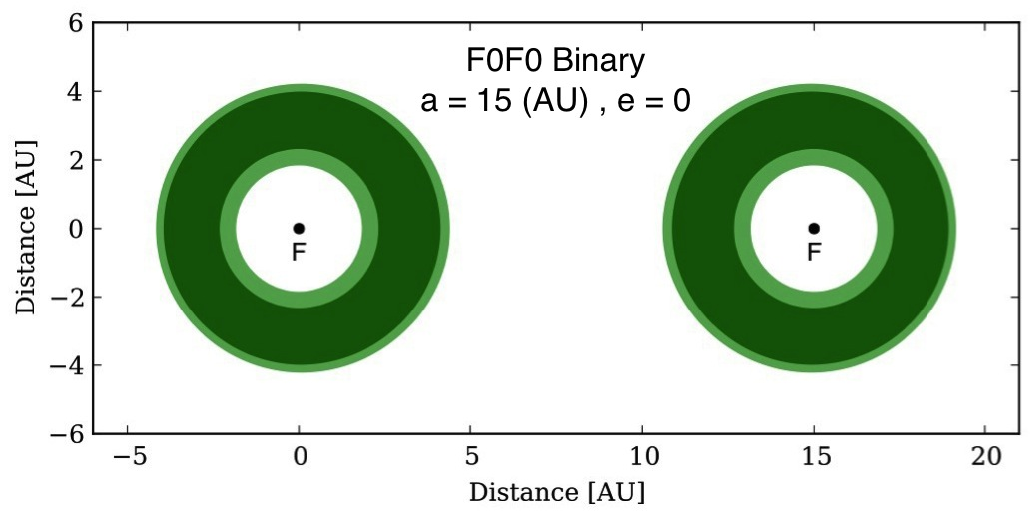}
}
\vskip 5pt
\makebox[\textwidth]{
\includegraphics[width=.5\columnwidth]{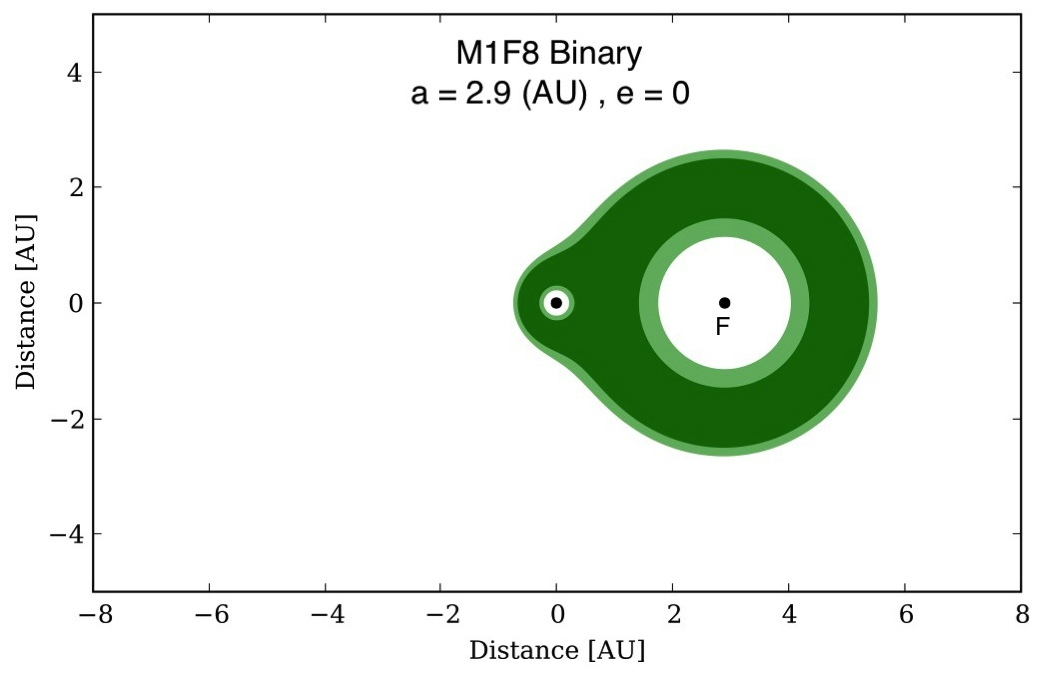}
\hfil
\includegraphics[width=.5\columnwidth]{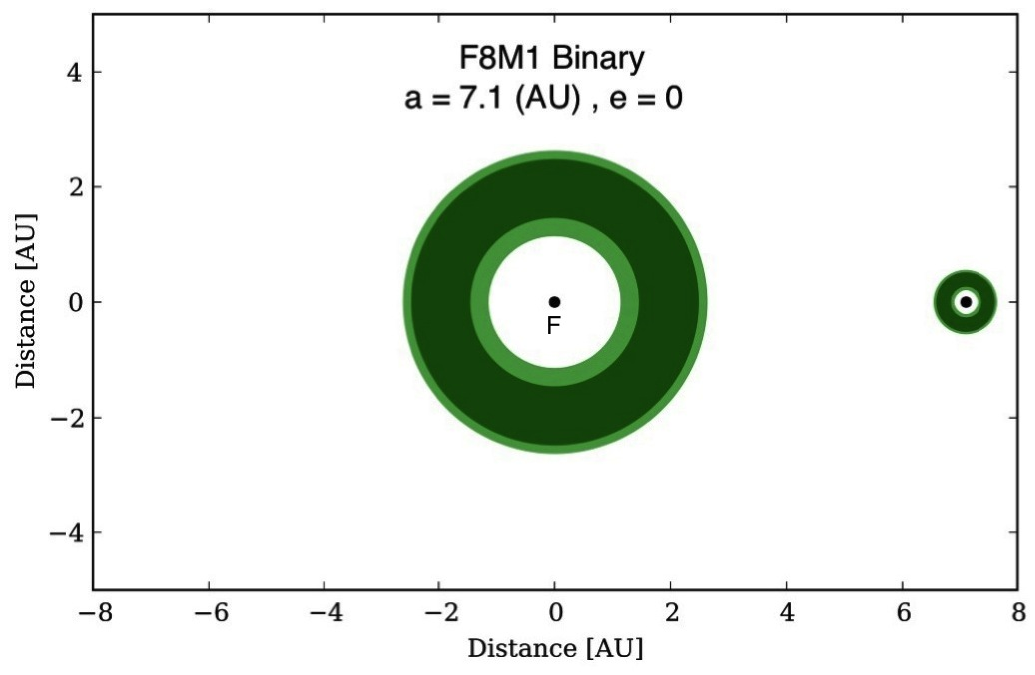}
}
\caption[]{Boundaries of the narrow (dark green) and empirical (light green) HZs in an M2-M2 (top left), F0-F0 (top right), and M1-F8 S-type binary star system (bottom two panels). Note that the primary is the star at (0,0). The primary star in the bottom panels is the F8 star (left) and the M1 star (right). The semimajor axis of the binary has been chosen to be the minimum value that allows the region out to the outer edge of the primary's empirical HZ to be stabile in a circular binary.
}

\label{Boundaries}
\end{figure*}

To explore the maximum effect of the binary semimajor axis and eccentricity on the contribution of one star to the extent of the HZ around the other component, we consider three extreme cases: an M2-M2, an F0-F0, and an F8-M1 binary. We consider the M2 and F0 stars to have effective temperatures of 3520 (K) and 7300 (k), respectively, and their luminosities to be 0.035 and 6.56 of that of the Sun for our general examples here. We note that in calculating the boundaries of the HZ, the orbital (in)stability of the fictitious Earth-like planet is not considered. As a result, depending on the orbital elements and mass-ratio of the binary, its HZ may be unstable.

\begin{figure*}
\makebox[\textwidth]{
\includegraphics[width=.5\columnwidth]{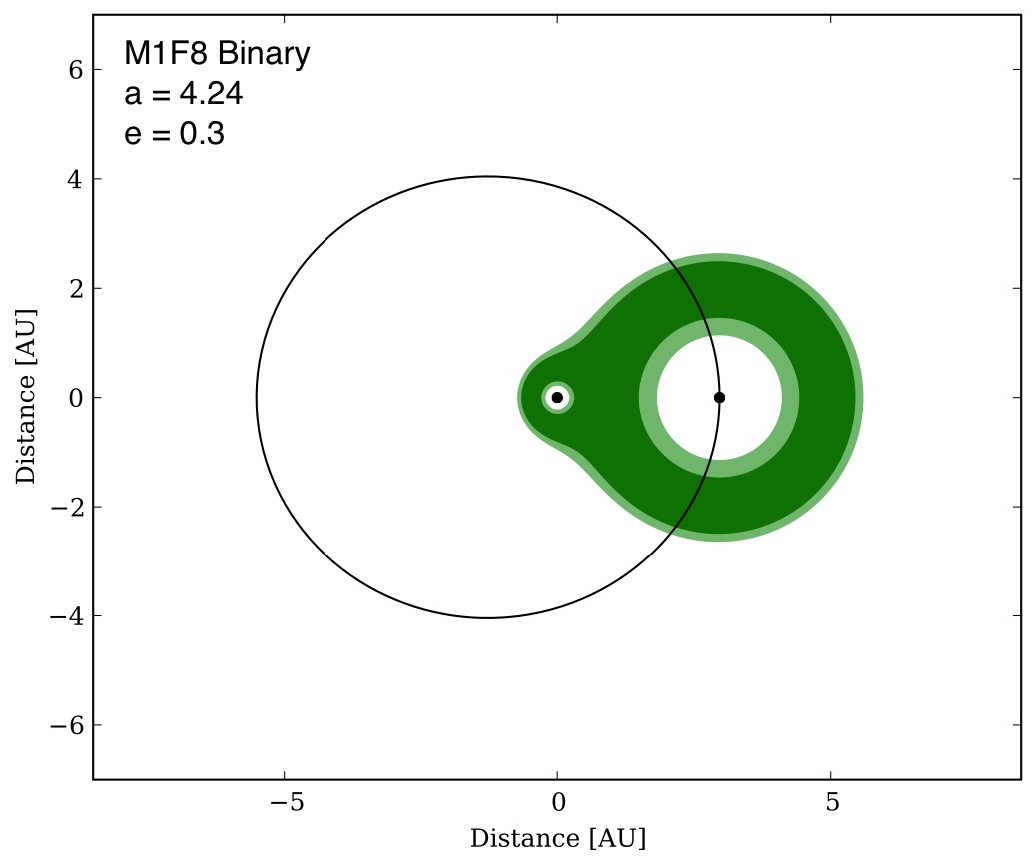}
\hfil
\includegraphics[width=.5\columnwidth]{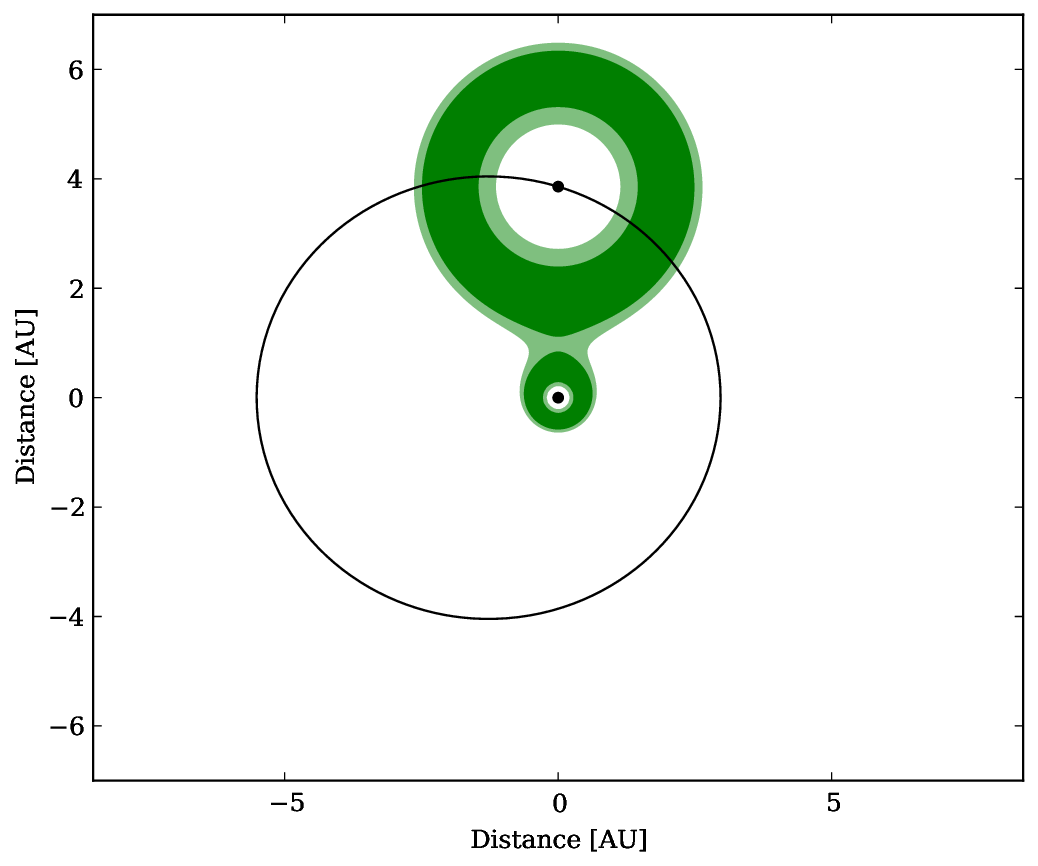}
}
\vskip 2in
\makebox[\textwidth]{
\includegraphics[width=.5\columnwidth]{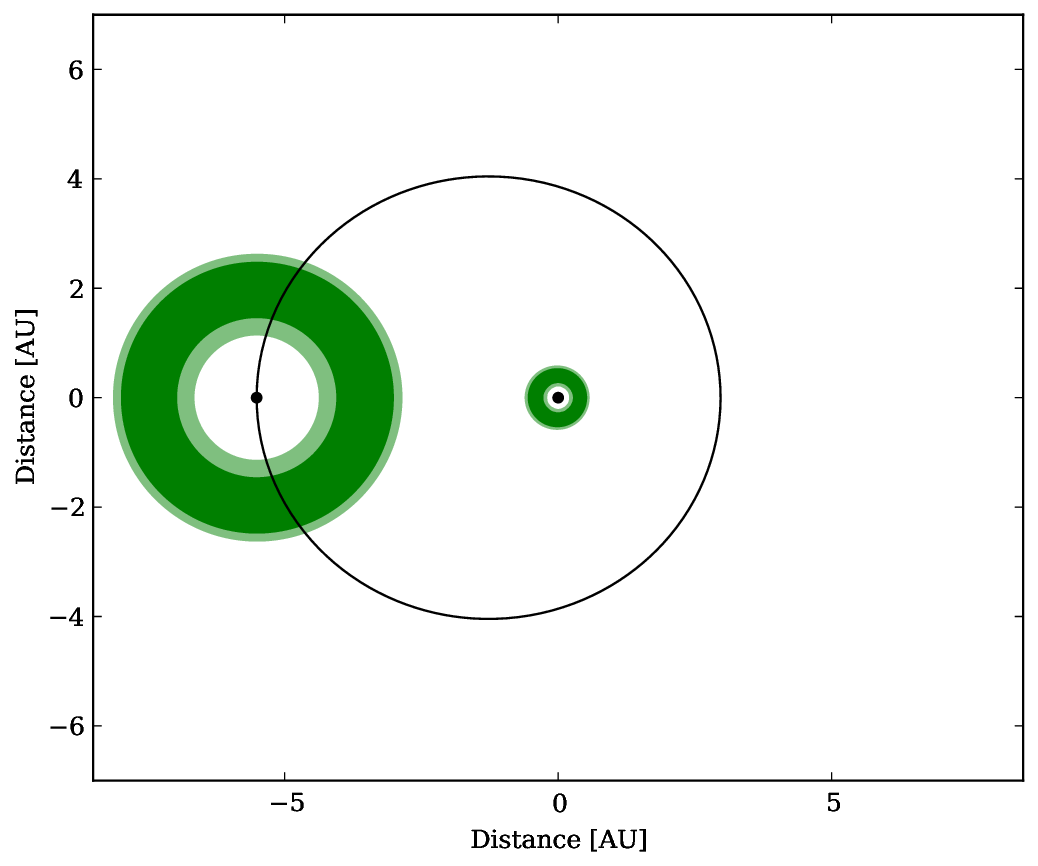}
\hfil
\includegraphics[width=.5\columnwidth]{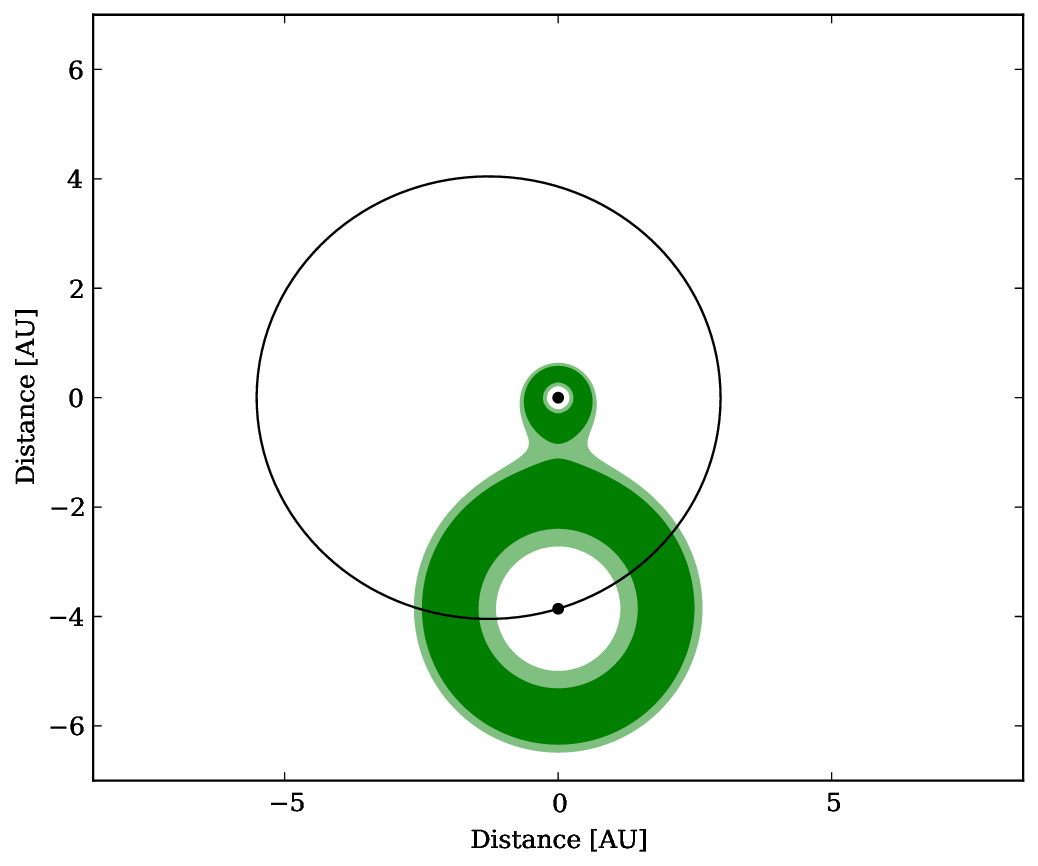}
}
\caption[]{Boundaries of the narrow (dark green) and empirical (light green) HZs in an M1-F8 binary. Note that the primary is the M1 star at (0,0). The panels show the effect of the F8 star while orbiting the primary starting from the top left panel when the secondary is at the binary periastron. The semimajor axis of the binary has been chosen to be the minimum value that allows the region out to the outer edge of the primary'€™s empirical single-star HZ to be stable for a binary eccentricity of 0.3.
}
\label{BoundMF}
\end{figure*}

To demonstrate the effect of the secondary on the boundaries of the HZ, we calculate the Binary HZ of the systems mentioned above considering the minimum value of the binary semimajor axis for which the outer edge of the primary's empirical HZ will be on the stability limit. Fig.\ref{BoundMF} shows the results for the case of a circular binary. The top panels in this figure correspond to an M2M2 (left) and F0F0 (right) binary system. As shown here, the secondary does not have a noticeable effect on the extent of the HZ around the primary. The Binary HZ around each star is equivalent to its single-star HZ. The bottom panels in Fig.\ref{BoundMF} correspond to an F8M1 binary (left) where the primary is the F star, and an M1F8 binary (right) where the primary is the M star. As expected, the effect of the M star on the extension of the single-star HZ around the F star is negligible. However, at its closets distances, the F star can extent the outer limit of the single-star HZ around the M star so far out that at the binary periastron, the two HZs merge.

To explore the effect of binary eccentricity in a system with a hot and cool star, we carried out similar calculations as those in figure 13, for the F8M1 binary, assuming the binary eccentricity to be 0.3. Figures 14 and 15 show the results for four different relative potions of the two stars. In figure 14, the primary is the M star and in figure 15, the primary is the F star. As shown in these figures, when the secondary is more luminous, it will have considerable effects on the shape and location of the single-star HZ around the other star. However, a cool and less luminous secondary will not change the limits of the primary's single-star HZ.

\begin{figure*}
\makebox[\textwidth]{
\includegraphics[width=.5\columnwidth]{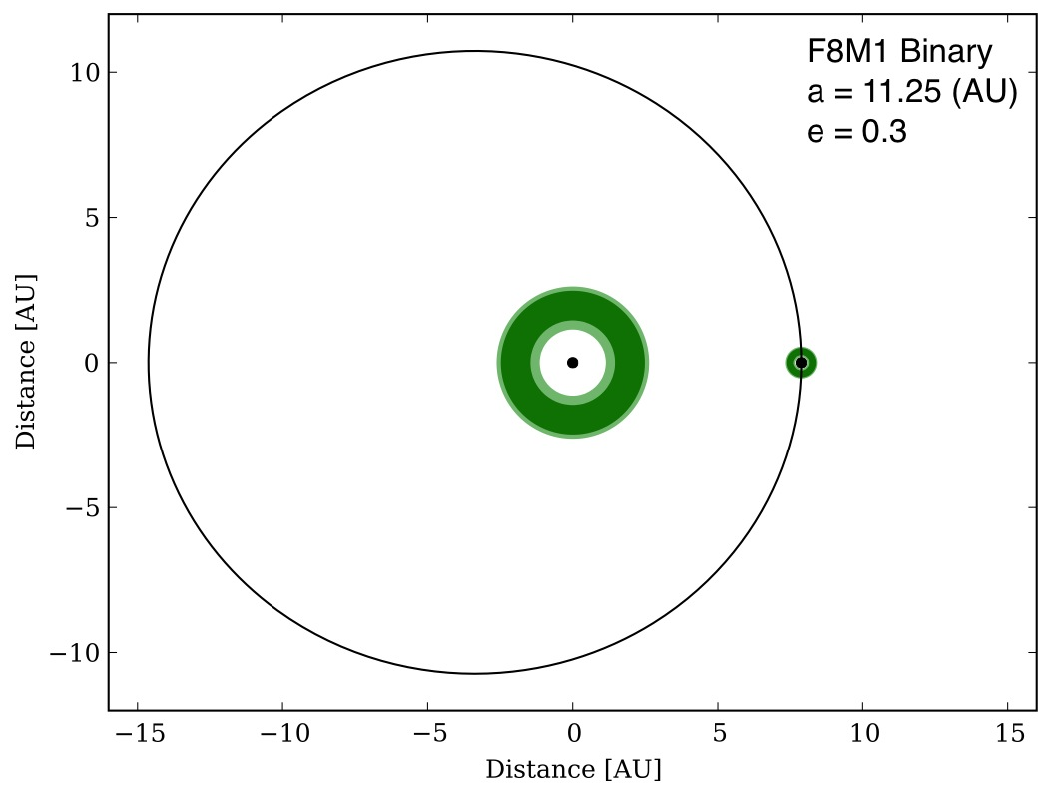}
\hfil
\includegraphics[width=.5\columnwidth]{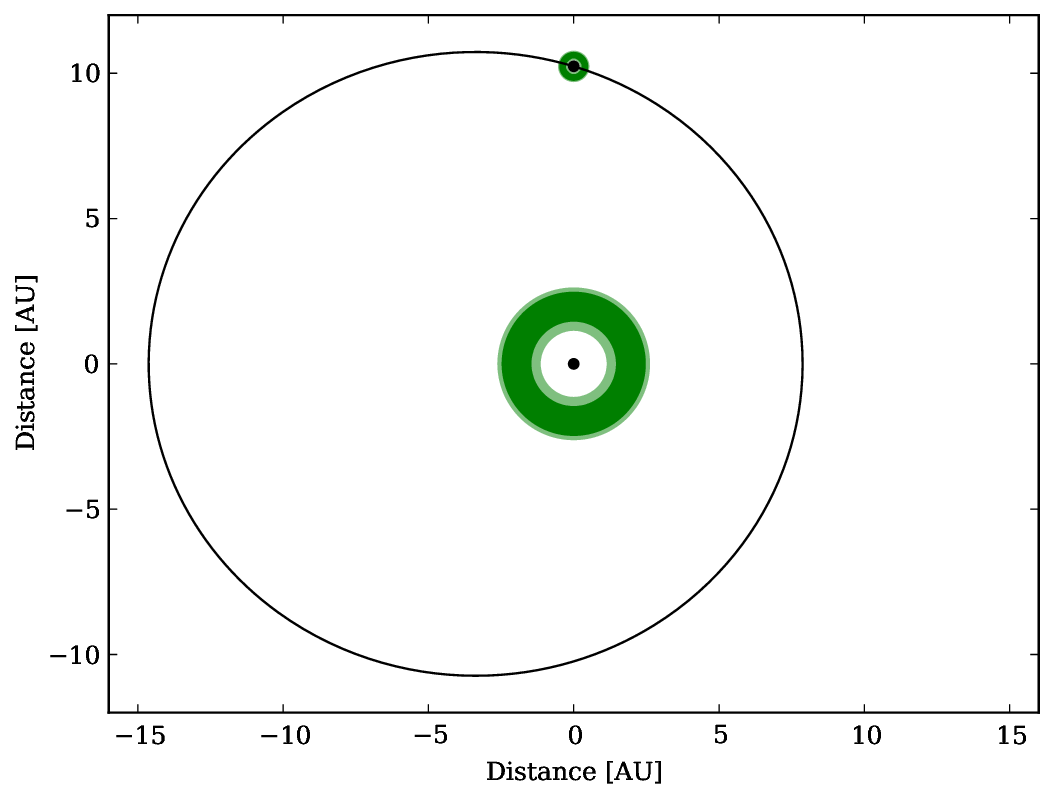}
}
\vskip 1.8in
\makebox[\textwidth]{
\includegraphics[width=.5\columnwidth]{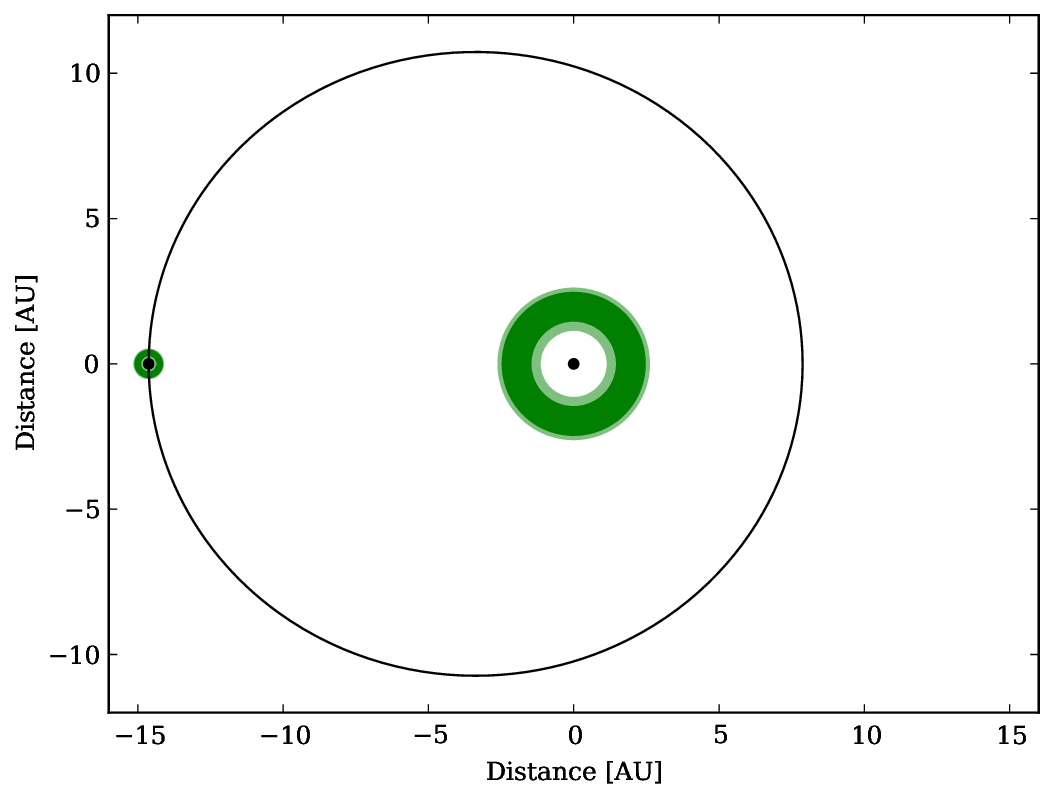}
\hfil
\includegraphics[width=.5\columnwidth]{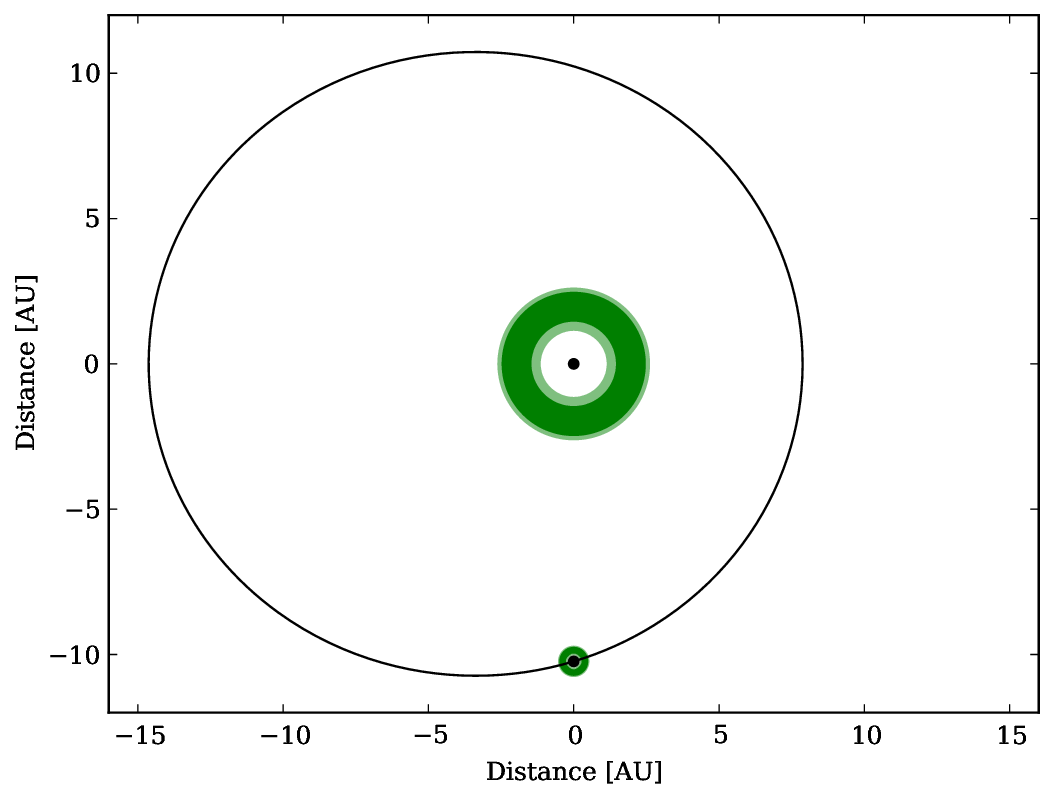}
}
\caption[]{Same as Figure 14, with the F star as the primary.}

\label{BoundF}
\end{figure*}

\subsection{Habitable Zone of $\alpha$ Centauri}

The $\alpha$ Centauri system consists of the close binary $\alpha$ Cen AB, and a farther M dwarf companion known as Proxima Centauri at approximately 15000 AU. The semimajor axis of the binary is $\sim 23.5$ AU and its eccentricity is $\sim 0.518$. The component A of this system is a G2V star with a mass of $1.1\, {M_{\rm Sun}}$, luminosity of $1.519\,{L_{\rm Sun}}$, and an effective temperature of 5790 K. Its component B has a spectral type of K1V, and its mass, luminosity and effective temperature are equal to $0.934\, {M_{\rm Sun}}$, $0.5\,{L_{\rm Sun}}$, and 5214 K, respectively.

The announcement of a probable super-Earth planet with a mass larger than 1.13 Earth-masses around $\alpha$ Cen B (Dumusque et al. 2012) indicates that unlike the region around $\alpha$ Cen A where terrestrial planet formation encounters complications, planet formation is efficient around this star (Guedes et al. 2008, Th\'ebault, Marzari \& Scholl 2009) and it could also host a terrestrial planet in its HZ. Here we assume that planet formation around both stars of this binary can proceed successfully and they both can host Earth-like planets. We calculate the spectral weight factors of both $\alpha$ Cen A and B (Table 2), and using the minimum and maximum added flux of the secondary star, estimate the limits of their Binary HZs using equation (2). 

Table 3 shows the estimates of the boundaries of the Binary HZ around each star. The terms Max and Min in this table correspond to the planet-binary configurations of $(\theta,\nu)=(0,0)$ and $(0,{180^\circ})$ where the planet receives the maximum and minimum total flux from the secondary star, respectively. As shown here, each star of the $\alpha$ Centauri system contributes to increasing the limits of the Binary HZ around the other star. Although these contributions are small, they extend the estimated limits of the Binary HZ by a noticeable amount. This is primarily due to the high luminosity of $\alpha$  Cen A and the relatively large eccentricity of the binary which brings the two stars as close as $\sim 11.3$ AU from one another (and as such making planet formation very difficult around $\alpha$ Cen A).

Given the eccentricity of the $\alpha$ Cen binary $({e_{\rm Bin}}=0.518)$, both narrow and nominal HZs for $\alpha$ Cen A and B are stable. The stability limit around the primary G2V star is at $\sim 2.768$ AU. This limit is slightly exterior to the outer boundary of the star's narrow and empirical HZs. Although the latter suggests that the HZ of $\alpha$ Cen A is dynamically stable, the close proximity of this region to the stability limit may have strong consequences on the actual formation of an Earth-like planet in this region (see e.g. Th\'ebault et al. 2008, Eggl 2012).

\subsection {Habitable Zone of HD 196885}

\begin{table}
\centering
\caption{Samples of Spectral Weight Factors}
\begin{tabular}{l|c|c|c|c|c}
    &  & \multicolumn{2}{c}{\underline{Narrow HZ}} & 
\multicolumn{2}{c}{\underline{Empirical HZ}} \\
 Star   &  Eff. Temp & inner & outer &  inner & outer \\
\hline
F0                   & 7300 & 0.850 & 0.815 & 0.902 & 0.806 \\
F8V (HD 196885 A)    & 6340 & 0.936 & 0.915 & 0.957 & 0.913 \\
G0                   & 5940 & 0.981 & 0.974 & 0.987 & 0.973 \\
G2V ($\alpha$ Cen A) & 5790 & 0.999 & 0.998 & 0.999 & 0.998 \\
K1V ($\alpha$ Cen B) & 5214 & 1.065 & 1.100 & 1.046 & 1.103 \\
K3                   & 4800 & 1.107 & 1.179 & 1.079 & 1.186 \\
M1V (HD 196885 B)    & 3700 & 1.177 & 1.383 & 1.154 & 1.419 \\
M5                   & 3170 & 1.192 & 1.471 & 1.179 & 1.532 \\
M2                   & 3520 & 1.183 & 1.414 & 1.163 & 1.458 \\
\end{tabular}
\end{table}

\begin{table}
\centering
\caption{Estimates of the boundaries of the Binary HZ calculated using equation (2) for the maximum and minimum amount of flux 
received by the planet from the secondary star corresponding to the closest and farthest distances between the two bodies.}
\begin{tabular}{l|c|c|c|c|c|c|c|c}
Host Star & \multicolumn{4}{c}{Estimates of Narrow HZ (AU)} & \multicolumn{4}{c}{Estimates of Empirical HZ (AU)} \\
\hline
   & \multicolumn{2}{c}{\underline{With Secondary}} &  \multicolumn{2}{c}{\underline{Without Secondary}} & 
\multicolumn{2}{c}{\underline{With Secondary}} &  \multicolumn{2}{c}{\underline{Without Secondary}} \\
    & inner & outer &  inner & outer  & inner & outer &  inner & outer  \\
\hline
$\alpha$ Cen A (Max) & 1.197 & 2.068 & 1.195 & 2.056 & 0.925 & 2.194 & 0.924 & 2.179\\
$\alpha$ Cen A (Min)  & 1.195 & 2.057 & 1.195 & 2.056 & 0.924 & 2.181 & 0.924 & 2.179    \\
\hline
$\alpha$ Cen B (Max) & 0.712 & 1.259 & 0.708 & 1.238 & 0.544 & 1.340 & 0.542 & 1.315\\
$\alpha$ Cen B (Min) & 0.708 & 1.241 & 0.708 & 1.238 & 0.543 & 1.317 & 0.542 & 1.315    \\
\hline
HD 196885 A (Max)  & 1.454 & 2.477 & 1.454 & 2.475 & 1.137 & 2.622 & 1.137 & 2.620 \\
HD 196885 A (Min)   & 1.454 & 2.475 & 1.454 & 2.475 & 1.137 & 2.620 & 1.137 & 2.620    \\
\hline
HD 196885 B (Max)  & 0.260 & 0.491 & 0.258 & 0.481 & 0.198 & 0.529 & 0.197 & 0.516  \\
HD 196885 B (Min)   & 0.258 & 0.483 & 0.258 & 0.481 & 0.197 & 0.518 & 0.197 & 0.516    \\
\end{tabular}
\end{table}

HD 196885 is a close main sequence S-type binary system with a semimajor axis of 21 AU and eccentricity of 0.42 (Chauvin et al. 2011).  The primary of this system (HD 169885 A) is an F8V star with a $T_{\rm Star}$ of 6340 K, mass of 1.33 ${M_{\rm Sun}}$, and luminosity of 2.4 $L_{\rm Sun}$. The secondary star (HD 196885 B) is an M1V dwarf with a mass of $0.45 {M_{\rm Sun}}$. Using the mass-luminosity relation $L \sim {M^{3.5}}$ where $L$ and $M$ are in solar units, the luminosity of this star is approximately $0.06 {L_{\rm Sun}}$, and we consider its effective temperature to be ${T_{\rm Star}}=3700$ K. The primary of HD 196885 hosts a Jovian-type planet suggesting that the mass-ratio and orbital elements of this binary allow planet formation to proceed successfully around its primary star. We assume that terrestrial planet formation can also successfully proceed around both stars of this binary and can result in the formation of Earth-sized objects. 

To estimate the boundaries of the Binary HZ of this eccentric system, we ignore its known giant planet and use equation (2) considering a fictitious Earth-like planet in the HZ.  We calculate the spectral weight factor $W(f,T)$ for both stars of this system (Table 2) and estimate the locations of the inner and outer boundaries of the binary's HZ (Table 3). 

As expected (because of the large periastron distance of the binary, and the secondary star being a cool M dwarf), even the maximum flux from the secondary star does not have a noticeable contribution to the location of the HZ around the F8V primary of HD 196886. However, being a luminous F star, the primary shows a small effect on the location of the HZ around the M1V secondary star (Table 3).


\begin{thebibliography}{}
%
\bibitem[Abe et al.(2011)]{abe11} Abe, Y., Abe-Ouchi, A., Sleep, N. H.,  Zahnle, K. J. 2011, AsBio, 11, 443
%
\bibitem[Artymowicz \& Lubow(1994)]{arty94} Artymowicz, P., Lubow, S. H., 1994, ApJ, 421, 621
%
\bibitem[Barbieri et al.(2002)]{barb02} Barbieri, M.; Marzari, F.; Scholl, H., 2002, Formation of terrestrial planets in close binary systems: The case of alpha Centauri A, A\&A, 396, 219
%
\bibitem[Beauge et al.(2010)]{beau10} Beauge, C., Leiva, A. M.; Haghighipour, N.; Otto, J. Correa, 2010, Dynamics of planetesimals due to gas drag from an eccentric precessing disc, MNRAS, 408, 503
%
\bibitem[Blum \& Wurm(2008)]{blum08} Blum, J\"urgen, Wurm, Gerhard, 2008, The Growth Mechanisms of Macroscopic Bodies in Protoplanetary Disks, ARA\&A, 46, 21
%
\bibitem[Boley(2009)]{bole09} Boley, A.C., 2009, The Two Modes of Gas Giant Planet Formation, ApJ, 695, L53
%
\bibitem[Boss(1997)]{boss97} Boss, A.P., 1997, Giant planet formation by gravitational instability, Science, 276, 1836
%
\bibitem[Boss(2011)]{boss11} Boss, A.P., 2011, Formation of Giant Planets by Disk Instability on Wide Orbits Around Protostars with Varied Masses, ApJ, 731, 74
%
\bibitem[Boss(2006)]{boss06} Boss, A.P., 2006, Gas Giant Protoplanets Formed by Disk Instability in Binary Star Systems, ApJ, 641, 1148
%
\bibitem[Chauvin et al.(2011)]{chau10} Chauvin, G.; Beust, H.; Lagrange, A. -M.; Eggenberger, A., 2011, Planetary systems in close binary stars: the case of HD 196885. Combined astrometric and radial velocity study, A\&A, 528, 8
%
\bibitem[Correia et al.(2008)]{corr08} Correia, A.C., and 11 co-authors, 2008, The ELODIE survey for northern extra-solar planets. IV. HD 196885, a close binary star with a 3.7-year planet, A\&A, 479, 271
%
\bibitem[Cuzzi et al.(2008)]{cuzz08} Cuzzi, Jeffrey N.; Hogan, Robert C.; Shariff, Karim, 2008, Toward Planetesimals: Dense Chondrule Clumps in the Protoplanetary Nebula, ApJ, 687, 1432
%
\bibitem[Desidera \& Barbieri(2007)]{desi07} Desidera, S., Barbieri, M.,
2007, Properties of planets in binary systems. The role of binary separation, A\&A 462, 345-353
%
\bibitem[Doyle et al.(2011)]{doyl11} Doyle, L., Carter, J., et al., 2011, Kepler-16: A Transiting Circumbinary Planet, Science, 333, 1602
%
\bibitem[Duch\^ene(2010)]{duch10} Duch\^ene, G., 2010, Planet Formation in Binary Systems: A Separation-Dependent Mechanism?, ApJ, 709, L114
%
\bibitem[Duch\^ene \& Kraus(2013)]{duch13} Duch\^ene, G., Kraus, A., 2013, ARA\&A, 51
%
\bibitem[Dumusque et al.(2012)]{dumu12} Dumusque, X., Pepe, F. et al., 2012, An Earth-mass planet orbiting alpha Centauri B, Nature, 491, 207
%
\bibitem[Duquennoy and Mayor(1991)]{duq91} Duquennoy, A.; Mayor, M., 1991, Multiplicity among solar-type stars in the solar neighbourhood. II - Distribution of the orbital elements in an unbiased sample, A\&A, 248, 485
%
\bibitem[Dvorak (1984)]{dvo84} Dvorak, R.; 1984, Numerical experiments on planetary orbits in double stars, CeMec, 34, 369
%
\bibitem[Dvorak (1986)]{dvo86} Dvorak, R.; 1986, Critical orbits in the elliptic restricted three-body problem, A\&A, 167, 379
%
\bibitem[Dvorak et al.(1989)]{dvo89} Dvorak, R., Froeschle, C.; Froeschle, Ch., 1989, Stability of outer planetary orbits (P-types) in binaries, A\&A, 226, 335
%
\bibitem[Dvorak et al.(2003)]{dvo03} Dvorak, R.; Pilat-Lohinger, E.; Funk, B.; Freistetter, F. , 2003, Planets in habitable zones:. A study of the binary Gamma Cephei, A\&A, 398, L1
%
\bibitem[Eggl et al.(2012)]{eggl12} Eggl, S., Pilat-Lohinger, E., Georgakarakos, N., Gyergyovits, M., Funk, B. 2012, ApJ, 752, 74
%
\bibitem[Eggl et al.(2013a)]{eggl13a} Eggl, S., Haghighipour, N., Pilat-Lohinger, E. 2013, ApJ, 764, 130
%
\bibitem[Eggl et al.(2013b)]{eggl13b} Eggl, S., Pilat-Lohinger, E., Funk, B., Georgakarakos, N.,  Haghighipour, N. 2013, MNRAS, 428, 3104
%
\bibitem[Eggenberger \& Udry(2010)]{egge10} Eggenberger, A., Udry, S., 2010, Probing the Impact of Stellar Duplicity on Planet Occurrence with Spectroscopic and Imaging Observations, in Planets in Binary Star Systems, ed. N.Haghighipour (New York: Springer), 19
%
\bibitem[Endl et al.(2011)]{endl11} Endl, M., Cochran, W., Hatzes, A., Wittenmeyer, R.A., 2011, News from the $\gamma$ Cephei Planetary System, PLANETARY SYSTEMS BEYOND THE MAIN SEQUENCE: Proceedings of the International Conference. AIP Conference Proceedings, 1331, 88.
%
\bibitem[Fragner et al.(2011)]{frag11} Fragner, M., Nelson, R., Kley, W., 2011, On the dynamics and collisional growth of planetesimals in misaligned binary systems A\&A, 528, 40
%
\bibitem[Georgakarakos(2002)]{geor02} Georgakarakos, N. 2002, MNRAS, 337, 559
%
\bibitem[Giuppone et al.(2011)]{giup11} Giuppone, C. A.; Leiva, A. M.; Correa-Otto, J.; Beaug\'e, C., 2011, Secular dynamics of planetesimals in tight binary systems: application to $\gamma$ Cephei, A\&A, 530, 103
%
\bibitem[Guedes et al.(2008)]{gued08} Guedes, J. M.; Rivera, E. J.; Davis, E.; Laughlin, G.; Quintana, E. V.; Fischer, D. A, 2008, Formation and Detectability of Terrestrial Planets around alpha Centauri B, ApJ, 679, 1582
%
\bibitem[Haghighipour (2006)]{hagh06}Haghighipour, N. 2006, ApJ, 644, 543
%
\bibitem[Haghighipour \& Raymond (2007)]{hagh07} Haghighipour, N., \&  Raymond, S. N. Habitable Planet Formation in Binary Planetary Systems, 2007, ApJ, 666, 436--446.
%
\bibitem[Haghighipour  et al.(2010)]{hagh10} Haghighipour, Nader; Dvorak, Rudolf; Pilat-Lohinger, Elke, 2010, Planetary Dynamics and Habitable Planet Formation in Binary Star Systems, ASSL, 366, 285
%
\bibitem[Haghighipour (2011)]{hagh11} Haghighipour N., 2011, Super-Earths: a new class of planetary bodies, Contemp. Phys. 52:403–38
%
\bibitem[Haghighipour (2013)]{hagh13} Haghighipour N., 2013, The Formation and Dynamics of Super-Earths, Annu. Rev. Earth Planet. Sci. 2013. 41:469–95
%
\bibitem[Hale(1994)]{hale94} Hale, A., 1994, Orbital coplanarity in solar-type binary systems: Implications for planetary system formation and detection, AJ, 107, 306
%
\bibitem[Harris et al.(2012)]{harr12} Harris, R. J.; Andrews, S. M.; Wilner, D. J.; Kraus, A. L., 2012, A Resolved Census of Millimeter Emission from Taurus Multiple Star Systems, ApJ, 751, 115
%
\bibitem[Hatzes et al.(2003)]{hatz03} Hatzes, A. P., Cochran, W. D., Endl, M., McArthur, B., Paulson, D. B., Walker, G.A.H., Campbell, B., Yang, S., 2003, A Planetary Companion to Gamma Cephei A, ApJ, 599, 1383
%
\bibitem[Hayashi(1981)]{haya81} Hayashi, C., 1981, Structure of the Solar Nebula, Growth and Decay of Magnetic Fields and Effects of Magnetic and Turbulent Viscosities on the Nebula, PthPS 70, 35  
%
\bibitem[Heppenheimer(1978)]{hepp78} Heppenheimer, T.A., 1978, On the formation of planets in binary star systems, A\&A, 65, 421
%
\bibitem[Holman \& Wiegert(1999)]{holw99} Holman, M.J., Wiegert, P. A., 1999, Long-Term Stability of Planets in Binary Systems, AJ, 117, 621
%
\bibitem[Hubickyj et al.(2005)]{hubi05} Hubickyj O, Bodenheimer P, Lissauer JJ. 2005. Accretion of the gaseous envelope of Jupiter around a 5–10 Earth-mass core. Icarus 179:415–31
%
\bibitem[Jang-Condell et al.(2008)]{jang08} Jang-Condell, H.; Mugrauer, M.; Schmidt, T., 2008, 
Disk Truncation and Planet Formation in Gamma Cephei, ApJ, 683, L191
%
\bibitem[Johansen et al.(2007)]{joha07} Johansen, Anders; Oishi, Jeffrey S.; Mac Low, Mordecai-Mark; Klahr, Hubert; Henning, Thomas; Youdin, Andrew, 2007, Rapid planetesimal formation in turbulent circumstellar disks, Nature, 448, 1022
%
\bibitem[Kaltenegger \& Sasselov(2011)]{kalt11} Kaltenegger, L., Sasselov, D. 2011, ApJL, 736, L25
%
\bibitem[Kasting et al.(1993)]{kast1993} Kasting, J. F., Whitmire, D. P.,  Reynolds, R. T. 1993, Icarus, 101, 108
%
\bibitem[Kley \& Nelson(2007)]{kley07} Kley, W., Nelson, R. P., 2007, in "Planets in binary  Star Systems," ed. Nader Haghighipour (Springer publishing company)
%
\bibitem[Kley \& Nelson(2008)]{kley08} Kley, W., Nelson, R. P., 2008, Planet formation in binary stars: the case of Gamma Cephei, A\&A, 486, 617
%
\bibitem[Kokubo \& Ida(2000)]{koku00} Kokubo, E., Ida, S., 2000, Formation of Protoplanets from Planetesimals in the Solar Nebula, Icarus, 297, 1067
%
\bibitem[Kopparapu et al.(2013a)]{kopp13a} Kopparapu, R. K., Ramirez, R., Kasting, J. F., et al. 2013a, ApJ, 765, 131
%
\bibitem[Kopparapu et al.(2013b)]{kopp13b} Kopparapu, R. K., Ramirez, R., Kasting, J. F., et al. 2013b, ApJ, 770, 82
%
\bibitem[kostov et al.(2013)]{kost13} Kostov, V. B., et al., 2013, ApJ, 770, 52
%
\bibitem[kostov et al.(2014a)]{kost14a} Kostov, V. B., et al., 2014, ApJ, 784, 14
%
\bibitem[kostov et al.(2014b)]{kost14b} Kostov, V. B., et al., 2014, ApJ, 787, 93
%
\bibitem[Kraus et al.(2012)]{krau12} Kraus, A. L.; Ireland, M. J.; Hillenbrand, L. A.; Martinache, F., 2012, The Role of Multiplicity in Disk Evolution and Planet Formation, ApJ, 745, 12
%
\bibitem[Lagrange et al.(2006)]{lag06} Lagrange, A.-M.; Beust, H.; Udry, S.; Chauvin, G.; Mayor, M., 2006, New constrains on Gliese 86 B. VLT near infrared coronographic imaging survey of planetary hosts, A\&A, 459, 955
%
\bibitem[Lissauer (1993)]{liss93} Lissauer, J.J., 1993, Planet formation, ARA\&A, 31, 129
%
\bibitem[Lodders \& fegley(1998)]{lodd98} Lodders, K., Fegley, B., Jr. 1998, Meteoritics Planet. Sci., 33, 871
%
\bibitem[Makino et al.(1998)]{maki98} Makino, J., Fukushige, T., Funato, Y., Kokubo, E., 1998, On the mass distribution of planetesimals in the early runaway stage, NewA, 3, 411
%
\bibitem[Malmberg et al.(2007)]{malm07} Malmberg, D.; Davies, M. B.;
Chambers, J. E., 2007, Close encounters in young stellar clusters: implications for planetary systems in the solar neighbourhood, MNRAS, 378, 1207
%
\bibitem[Marzari \& Scholl(2000)]{marz00} Marzari, F., Scholl, H, 2000, 
Planetesimal Accretion in Binary Star Systems, ApJ, 543, 328
%
\bibitem[Marzari et al.(2005)]{marz05} Marzari, F., Weidenschilling, S.J., Barbieri, M., Granata, V., 2005, Jumping Jupiters in Binary Star Systems, ApJ, 618, 502
%
\bibitem[Marzari et al.(2008)]{marz08} Marzari, F., Thebault, P., Scholl, H, 2008, Planetesimal Evolution in Circumbinary Gaseous Disks: A Hybrid Model, ApJ, 681, 1599
%
\bibitem[Marzari et al.(2009)]{marz09} Marzari, F., Scholl, H., Thebault, P., Baruteau, C., 2009, On the eccentricity of self-gravitating circumstellar disks in eccentric binary systems, A\&A 508, 1493
%
\bibitem[Marzari et al.(2012)]{marz12} Marzari, Baruteau, C., Scholl, H., Thebault, P., 2012, Eccentricity of radiative disks in close binary-star systems, A\&A, 539, 98
%
\bibitem[Masset \& Snellgrove(2001)]{mass01} Masset, F., 2001, Reversing type II migration: resonance trapping of a lighter giant protoplanet, MNRAS, 320, 55
%
\bibitem[Mayer et al.(2005)]{maye05} Mayer, Lucio; Wadsley, James; Quinn, Thomas; Stadel, Joachim, 2005, Gravitational instability in binary protoplanetary discs: new constraints on giant planet formation, MNRAS, 363, 641
%
\bibitem[Mayer et al.(2010)]{maye10} Mayer, Lucio; Boss, Alan; Nelson, Andrew F., Gravitational Instability in Binary Protoplanetary Disks, In Planets in Binary Star Systems, ed. N. Haghighipour, Astrophysics and Space Science Library,
Volume 366. ISBN 978-90-481-8686-0. Springer
%
\bibitem[Minton \& Malhotra(2010)]{mint10} Minton, David A.; Malhotra, Renu, 2010, Dynamical erosion of the asteroid belt and implications for large impacts in the inner Solar System, Icarus, 207, 744
%
\bibitem[Movshovitz et al.(2010)]{movs10} Movshovitz N, Bodenheimer P, Podolak M, Lissauer JJ. 2010. Formation of Jupiter using opacities based on detailed grain physics. Icarus 209:616–24
%
\bibitem[Movshovitz \& Podolak(2008)]{movs08} Movshovitz N, Podolak M. 2008. The opacity of grains in protoplanetary atmospheres. Icarus 194:368–78
%
\bibitem[Mugrauer \& Neuh\"auser(2009)]{mugr09} Mugrauer, M.; Neuh\"auser, R., 2009, 
The multiplicity of exoplanet host stars. New low-mass stellar companions of the exoplanet host stars HD 125612 and HD 212301, A\&A, 494, 373
%
\bibitem[M\"uller \& Kley(2012)]{mull12} M\"uller, T.W.A., Kley, W., 2012, Circumstellar disks in binary star systems. Models for $\gamma$ Cephei and $\alpha$ Centauri, A\&A, 539, 18
%
\bibitem[Nelson(2000)]{nels00} Nelson, Andrew, 2000, Planet Formation is Unlikely in Equal-Mass Binary Systems with a = 50 AU, ApJ, 537, 65
%
\bibitem[Neuh\"auser et al.(2007)]{neuh07} Neuh\"auser, R.; Mugrauer, M.; Fukagawa, M.; Torres, G.; Schmidt, T., 2007, Direct detection of exoplanet host star companion Gamma Cep B and revised masses for both stars and the sub-stellar object, A\&A, 462, 777
%
\bibitem[Orosz et al.(2012a)]{oros12a} Orosz, J.A., Welsh, W.F., et al., 2012, Kepler-47: A Transiting Circumbinary Multiplanet System, Science, 337, 1511
%
\bibitem[Orosz et al.(2012b)]{oros12b} Orosz, J.A., Welsh, W.F., et al., 2012, The Neptune-sized Circumbinary Planet Kepler-38b, ApJ, 758, 87
%
\bibitem[Paardekooper et al.(2008)]{paard08} Paardekooper, S.-J., Thebault, P., \& Mellema, G., 2008, Planetesimal and gas dynamics in binaries, MNRAS, 386, 973 
%
\bibitem[Paardekooper \& Leinhardt(2010)]{paard10} Paardekooper, S.-J., Leinhardt, Z.M., 2010, Planetesimal collisions in binary systems,MNRAS, 403, L64
%
\bibitem[Payne et al.(2009)]{payn09} Payne, M.J., Wyatt, M.C., Thebault, P., 2009, Outward migration of terrestrial embryos in binary systems, MNRAS, 400, 1936
%
\bibitem[Picogna \& Marzari(2013)]{pico13} Picogna, G., Marzari, F., 2013, Three-dimensional modeling of radiative disks in binaries, A\&A, 556, 148
%
\bibitem[Pierens et al.(2012)]{pier12} Pierens, A., Baruteau, C., Hersant, F., 2012, Protoplanetary migration in non-isothermal discs with turbulence driven by stochastic forcing, MNRAS, 427, 1562
%
\bibitem[Pierrehumbert \& Gaidos(2011)]{pierr11} Pierrehumbert, R., Gaidos, E. 2011, ApJL, 734, L13
%
\bibitem[Plavchan et al.(2009)]{plav09} Plavchan, P.; Werner, M.W.; Chen, C.H.; Stapelfeldt, K.R.; Su, K.Y. L.; Stauffer, J.R.; Song, I., 2009, New Debris Disks Around Young, Low-Mass Stars Discovered with the Spitzer Space Telescope, ApJ, 698, 1068
%
\bibitem[Pollack et al.(1996)]{poll96} Pollack JB,Hubickyj O, Bodenheimer P, Lissauer JJ, Podolak M, Greenzweig Y, 1996, Formation of the giant planets by concurrent accretion of solids and gas, Icarus 124:62–85
%
\bibitem[Queloz et al.(2000)]{quel00} Queloz, D.; Mayor, M.; Weber, L.; Bl\'echa, A.; Burnet, M.; Confino, B.; Naef, D.; Pepe, F.; Santos, N.; Udry, S., 2000, The CORALIE survey for southern extra-solar planets. I. A planet orbiting the star Gliese 86, A\&A, 354, 99
%
\bibitem[Quian et al.(2010)]{qian10} Quian, S.-B., Lia, W.-P., Zhu, L.-Y., Dai, Z.-B., 2010, Detection of a Giant Extrasolar Planet Orbiting the Eclipsing Polar DP Leo, 708, 66
%
\bibitem[Quintana et al.(2002)]{quin02} Quintana, Elisa V.; Lissauer, Jack J.; Chambers, John E.; Duncan, Martin J, 2002, ApJ, 576, 982
%
\bibitem[Quintana et al.(2007)]{quin07} Quintana, E.~V., Adams, F.~C., Lissauer, J.J., Chambers, J.~E.\ 2007, ApJ, 660, 807 
%
\bibitem[Rafikov(2013)]{rafi13} Rafikov, R., 2013, Planet Formation in Small Separation Binaries: Not so Secularly Excited by the Companion, ApJ, 768, 112
%
\bibitem[Raghavan et al.(2010)]{ragh10} Raghavan, D.; McAlister, H. A.; Henry, T. J.; Latham, D. W.; Marcy, G. W.; Mason, B. D.; Gies, D. R.; White, R. J.; ten Brummelaar, T. A., 2010, A Survey of Stellar Families: Multiplicity of Solar-type Stars , ApJS, 190, 1
%
\bibitem[Roell et al.(2012)]{roel12} Roell, T., Neuh\"auser, R., Seifahrt, A., Mugrauer, M., Extrasolar planets in stellar multiple systems, A\&A, 542, 92
%
\bibitem[Rodriguez et al.(1998)]{rodr98} Rodr\'iguez, L. F.; D'Alessio, P.; Wilner, D. J.; Ho, P. T. P.; Torrelles, J. M.; Curiel, S.; G\'omez, Y.; Lizano, S.; Pedlar, A.; Cant\'o, J.; Raga, A. C., 1998, Compact protoplanetary disks around the stars of a young binary system, Nature, 395, 355
%
\bibitem[Safronov(1972)]{safr72} Safronov, V. S. 1972, Evolution of the protoplanetary cloud and formation of the earth and planets. (NASA-TTF-677)
%
\bibitem[Savonije et al.(1994)]{savo94} Savonije, G. J.; Papaloizou, J. C. B.; Lin, D. N. C., 1994, On Tidally Induced Shocks in Accretion Discs in Close Binary Systems, MNRAS, 268, 13
%
\bibitem[Schwamb et al.(2013)]{schw13} Schwamb, M. E., Orosz, J. A., Carter, J. A., et al., 2013, ApJ, 768, 127
%
\bibitem[Selsis et al.(2007)]{sels07} Selsis, F., Kasting, J. F., Levrard, B., et al. 2007, A\&A, 476, 1373
%
\bibitem[Stewart \& Leinhardt(2009)]{stew09} Stewart, Sarah T.; Leinhardt, Zoe M., 2009, Velocity-Dependent Catastrophic Disruption Criteria for Planetesimals, ApJ, 691, L133
%
\bibitem[Teiser \& Wurm(2009)]{teis09} Teiser, J.W., Wurm, G., 2009, High-velocity dust collisions: forming planetesimals in a fragmentation cascade with final accretion, MNRAS, 393, 1584
%
\bibitem[Thebault(2011)]{theb11} Thebault, P., 2011, Against all odds? Forming the planet of the HD 196885 binary, CeMDA, 111, 29
%
\bibitem[Thebault et al.(2006)]{theb06} Thebault, P., Marzari, F., Scholl, H., 2006, Relative velocities among accreting planetesimals in binary systems: The circumprimary case, Icarus, 183, 193
%
\bibitem[Thebault et al.(2004)]{theb04} Thebault, P., Marzari, F., Scholl, H., Turrini, D., Barbieri, M., 2004, Planetary formation in the Gamma Cephei system, A\&A, 427, 1097
%
\bibitem[Thebault et al.(2008)]{theb08} Thebault, P., Marzari, F., Scholl, H., 2008, Planet formation in alpha Centauri A revisited: not so accretion friendly after all, MNRAS, 388, 1528
%
\bibitem[Thebault et al.(2009)]{theb09} Thebault, P., Marzari, F., Scholl, H., 2009, Planet formation in the habitable zone of alpha Centauri B, MNRAS, 393, L21
%
\bibitem[Weidenschilling \& Davis(1985)]{wd85} Weidenschilling, S., Davis, D. R., 1985, Orbital resonances in the solar nebula - Implications for planetary accretion. Icarus, 62, 16 
%
\bibitem[Welsh et al.(2012)]{wels12} Welsh, W.F., Orosz, J.A., et al., Transiting circumbinary planets Kepler-34 b and Kepler-35 b, Nature, 481, 475
%
\bibitem[Xie \& Zhou(2008)]{xie08} Xie, Ji-Wei; Zhou, Ji-Lin, 2008, Planetesimal Accretion in Binary Systems: The Effects of Gas Dissipation, ApJ, 686, 570
%
\bibitem[Xie \& Zhou(2009)]{xie09} Xie, Ji-Wei; Zhou, Ji-Lin, 2009, Planetesimal Accretion in Binary Systems: Role of the Companion's Orbital Inclination
, ApJ, 698, 2066
%
\bibitem[Xie et al.(2010a)]{xie10} Xie, Ji-Wei; Zhou, Ji-Lin, Ge, Jian, 2010, Planetesimal Accretion in Binary Systems: Could Planets Form Around alpha Centauri B?, ApJ, 708, 1566
%
\bibitem[Xie et al.(2010b)]{xie10b} Xie, Ji-Wei; Payne, Matthew J.; Thebault, P.; Zhou, Ji-Lin; Ge, Jian, 2010, From Dust to Planetesimal: The Snowball Phase?, ApJ, 724, 1153
%
\bibitem[Zsom et al.(2011)]{zsom11} Zsom, Andras; Sandor, Zsolt; Dullemond, Cornelis, 2011, The first stages of planet formation in binary systems: how far can dust coagulation proceed?, A\&A, 527, 10
%
\bibitem[Zucker et al.(2004)]{zuck04} Zucker, S.; Mazeh, T.; Santos, N. C.; Udry, S.; Mayor, M., 2004, Multi-order TODCOR: Application to observations taken with the CORALIE echelle spectrograph. II. A planet in the system HD 41004, A\&A, 426, 695
%
\end{thebibliography}
\end{document}